\documentclass[aps,prd,reprint,superscriptaddress, floatfix]{revtex4-2}
\usepackage{color}
\usepackage{amssymb}
\usepackage{enumitem}   
\usepackage{graphicx,color}
\usepackage[usenames,dvipsnames]{xcolor}
\usepackage[section]{placeins}
\usepackage{amsmath}
\usepackage{subfigure}
\usepackage[normalem]{ulem}
\usepackage{booktabs}
\usepackage{xcolor}
\usepackage{epsfig,graphics,color,graphicx,amsmath}

\colorlet{darkgreen}{green!50!black}
\colorlet{brightyellow}{yellow!75!red}
\colorlet{orange}{red!50!yellow}
\colorlet{darkblue}{blue!60!black}
\colorlet{darkred}{red!80!black}

\def\be{\begin{eqnarray} &&}

\def\ee{\end{eqnarray}}

\usepackage{soul}

\usepackage{mathtools}
\usepackage{mathrsfs}
\usepackage{bm}
\usepackage{relsize}
\usepackage{indentfirst}

\usepackage{xcolor}

\begin{document}

\title{Proton image and momentum  distributions from light-front dynamics}

\author{E.~Ydrefors}
\email[]{ydrefors@kth.se}
\affiliation{Instituto Tecnol\'ogico de Aeron\'autica,  DCTA, 12228-900 S\~ao Jos\'e dos Campos,~Brazil}
\affiliation{Institute of Modern Physics, Chinese Academy of Sciences, Lanzhou 730000, China}
\author{T.~Frederico}
\email[]{tobias@ita.br}
\affiliation{Instituto Tecnol\'ogico de Aeron\'autica,  DCTA, 12228-900 S\~ao Jos\'e dos Campos,~Brazil}

\date{\today}

\begin{abstract}
We apply a dynamical three-constituent quark light-front model to study the proton. The dynamics is based on  the notion of a diquark  (bound or virtual) as the dominant interaction channel, which  parameterize a contact interaction between the quarks in order  to build the three-body Faddeev Bethe-Salpeter equations for the valence state, and we focus on the totally symmetric part of the wave function. 
The Dirac electromagnetic form factor is used to fix the model parameters, and the valence wave function is obtained. From that we investigate its Ioffe-time image, non-polarized longitudinal and transverse momentum distributions, and the double momentum distribution.
\end{abstract}
\maketitle

\section{Introduction}\label{Sec:intr}

The complex nucleon wave function on the null-plane $(x^+=t+z=0)$ expressed in the Fock space in terms of its constituent degrees of freedom, namely  quarks and gluons at a given scale $\mu$ and strongly interacting, ultimately provides the image through the associated probability densities~\cite{Brodsky:1997de,BakkerNPB2014,Arrington_2021}.
The relevant degrees of freedom at the hadronic scale are the dressed constituents, which carry the complex infrared (IR) physics,  namely, the confinement and  spontaneous chiral symmetry breaking ($\chi$SB). In particular, the dynamical $\chi$SB is reflected in the  large dressing of the light-flavored quarks, and also in the nucleon mass~\cite{Arrington_2021}.
It is also well known the large IR dressing of the gluon as computed with lattice QCD  (for a recent discussion of the gluon propagator in the Landau gauge see~\cite{li2020generalised}). 

The light-front (LF) wave function  is an eigenstate of the mass squared operator and the compatible operators $\vec P$ (momentum), $J^2$ (squared angular momentum), $J_z$,  and other compatible operators like the parity. However, $P_z$, $\vec J_\perp$, and parity are not diagonal in the Fock space, i.e.~they contain the interaction~\cite{Brodsky:1997de,BakkerNPB2014}. Although simple to state, still today the connection between QCD in Euclidean space, their infrared (IR) properties, with the LF wave function, and its Fock decomposition is yet a challenge for our understanding, beyond the large momentum behavior that thanks to the asymptotic freedom is well known, like the counting rules~(see e.g.~\cite{Brodsky:1997de})  as well as the ultraviolet (UV) behavior of the Fock amplitudes~\cite{Ji:2003fw}.

Ideally the LF nucleon wave function onto the null-plane should have an infinite number of Fock-state components that evolves with the renormalization scale $\mu$. The wave function 
can be decomposed into  Fock components each one
associated with a probability amplitude $\Psi_n(x_1,\vec k_{1\perp},x_2,\vec k_{2\perp},...;\mu)$ 
for $n\geq 3$ partons, which is invariant under  LF kinematical boosts. The probability corresponding to each Fock component is given by: 
\begin{equation}
\begin{aligned}
     P_n(\mu)=\Bigg\{
    \prod_{i=1 }^n\int\frac{d^2k_{i\perp}}{(2\pi)^2} \int^1_0 dx_i\Bigg\}\, \delta\left(1-\sum_{i=1}^n x_i\right)\, \\ \times\delta\left(\sum_{i=1}^n\vec k_{i\perp}\right) |\Psi_n(x_1,\vec k_{1\perp},x_2,\vec k_{2\perp},...;\mu)|^2\, ,
\end{aligned}
\end{equation}
where the transverse momentum of the constituent is $\vec k_{i\perp}$ and its longitudinal  momentum fraction  $x_i$.  We observe that  the probability $P_n(\mu)$ is invariant under LF kinematical boosts including translations on the null-plane hypersurface, so that we can choose, in particular, the frame where the transverse momentum vanishes.

The valence component corresponds to $n=3$. 
For simplicity, we have not depicted the dependence on the polarization state of the nucleon, as well as its constituents. The probability for each Fock component is $P_n(\mu)$, where we kept the scale dependence. At the hadron scale  $(\sim \Lambda_{QCD})$  the dominant component is the valence one, and for example in the pion case it
amounts to about 70\% as it has been computed recently in a Bethe-Salpeter (BS) framework~\cite{dePaula:2020qna}. The total normalization is
$  \sum_{i\geq 3}^ \infty P_n(\mu)=1$.

Each Fock amplitude can be written in the configuration space associated with the null-plane,
where the three-dimensional position coordinates for each constituent are 
$\{b^-_i,\vec b_{i\perp}\}$, namely the light-like coordinate $(b^-_i=t-z)$ and transverse position $(\vec b_{i\perp})$, conjugate to $k^+_i=x_ip^+$ and $\vec k_{i\perp}$, respectively. 
The  Fock component of the wave function on the null-plane is obtained by a Fourier transform:
\begin{small}
\begin{equation}
\begin{aligned}
 &   \tilde \Psi_n(\tilde x_1,\vec b_{1\perp},...;\mu)=\\
&    \Bigg\{\prod_{i=1 }^n\int\frac{d^2k_{i\perp}}{(2\pi)^2} \int^1_0 \frac{dx_i}{2\pi} \text{e}^{i\tilde{x}_i x_i-i\vec b_{i\perp}\cdot\vec k_{i\perp}} \Bigg\}
\\
&\times    \, \delta\left(1-\sum_{i=1}^n x_i\right)\, \delta\left(\sum_{i=1}^n\vec k_{i\perp}\right) \Psi_n(x_1,\vec k_{1\perp}...;\mu)\, ,
\end{aligned}
\end{equation}
\end{small}
where the dependence on the Ioffe time $\tilde x_i=b^-_ip^+$ ~\cite{Gribov,IoffePLB1969,Braun:1994jq} was given in the probability amplitude instead of the light-like position onto the null-plane.
The probability densities $|\tilde \Psi_n(\tilde x_1,\vec b_{1\perp},...;\mu)|^ 2$ build an image of the nucleon on the null-plane, where the light-like coordinate, shows the  relevance of the Ioffe time to complete the image of the nucleon (see a recent general discussion in~\cite{Miller:2019ysh} and in the case of the pion in ~\cite{dePaula:2020qna}).
Importantly, the Ioffe-time is a Lorentz invariant quantity which is related to the spatial distance between the struck quark and the spectators. Using the Ioffe-time one can through the inverse Fourier transform construct frame-independent PDFs. It should also be noted that the Ioffe-time representation of the PDF can be related at small space-like separations to the so-called Ioffe-time pseudo-distribution~\cite{Radyushkin:2017cyf}, which has been used to obtain the parton distribution of the pion from Lattice QCD~\cite{Joo:2019bzr}.

Different parton probability densities, namely one-, two- and N-body ones  can be defined given the LF wave function and reveal the multifaceted structure of the nucleon, which are associated with different observables being of interest not only for the present hadron facilities but also for the physics cases of the future Electron Ion Collider~\cite{khalek2021science}. In particular, we  mention the electromagnetic ones, as for example  the elastic form factor, and the parton distributions which are associated with one-body probability densities.

Other quantities  which can be computed from the LF wave function  are the generalized parton distributions (GPDs), being the non-perturbative objects entering the cross sections for deeply virtual Compton scattering (DVCS),  and the transverse momentum distributions  associated with semi-inclusive deeply inelastic scattering (SIDIS)~\cite{DiePRep03}. The PDFs  extracted from inclusive deep inelastic scattering give only information about the longitudinal momentum  fraction of the parton, i.e. simply  a one-dimensional view of the hadron. The  GPDs and  transverse momentum distributions (TMDs) provide a more complete image of the hadronic structure, in particular regarding the distribution of spin and orbital momentum in hadrons. That also allows a three-dimensional nucleon tomography in mixed position-momentum space~(see e.g. \cite{Lorce_2013}).
However, the most complete image is obtained through the six-dimensional Wigner distributions, and their Fourier transforms that are related to the generalized TMDs (GTMDs), which  can  appear in  the representation of hard  QCD processes~\cite{Meissner_2008,Meissner:2009ww,Lorce_2011,Lorce_2013}.  GTMDs are associated to  matrix elements of bi-local partonic  field operators with separation in all three light-front coordinates defined onto the null-plane hypersurface. In general, they are off-forward matrix elements between hadron states, which  depend on the partons longitudinal transverse momentum components. In particular,
the GTMDs correlate hadronic states with the same parton longitudinal momentum, namely for vanishing skewness, and different relative transverse distance between the struck partons' initial and final states.   It is worth to mention that, the GTMDs contain the bi-local correlators that define both the GPDs, TMDs, PDFs and as well as the electromagnetic form factor, which are obtained by taking certain limits or performing integrations, see e.g.~\cite{Lorce_2013}.

In particular, we remind that the space-like electromagnetic form factor can be obtained from the celebrated Drell-Yan-West formula~\cite{Brodsky:1997de} using the "plus" component of the current, with  momentum $q^+=0$ and $q^2=-\vec q_\perp^ 2 =-Q^2$, which is diagonal in the Fock space:
\begin{small}
\begin{equation}
  \label{Eq:G_EFS}
 \begin{aligned}
  F(&Q^2)=
  \sum_{n=3}^\infty F_{(n)}(Q^2)
  =\sum_{n=3}^\infty \sum_{j=1}^n e_j
  \\&\times\Bigg\{ \prod_{i=1 }^n\int\frac{d^2k_{i\perp}}{(2\pi)^2} \int^1_0 dx_i\Bigg\}\, \delta\left(1-\sum_{i=1}^n x_i\right)\, \\&\times\delta\left(\sum_{i=1}^n\vec k^\text{f}_{i\perp}\right)  \Psi_n^\dagger(x_1,\vec k^\text{f}_{1\perp},x_2,\vec k^\text{f}_{2\perp},...,x_j,\vec k^\text{f}_{j\perp},...)
\\ & \times    \Psi_n(x_1,\vec k^ \text{i}_{1\perp},x_2,\vec k^\text{i}_{2\perp},...,x_j,\vec k^\text{i}_{j\perp},...)\, ,
\end{aligned}
\end{equation}
\end{small}
where the number of constituents in the Fock components are $n\geq 3$, and $e_j$ is the constituent charge in units of the fundamental charge. The partial contribution to the form factor from each Fock component of the wave function is  $F_{(n)}(Q^2)$.  In the adopted frame the pair-creation contribution to the plus component of the current are suppressed, which is important since the present model of the proton is limited to the valence component, i.e.~$n=3$ in \eqref{Eq:G_EFS}.
The quark momenta obtained via the LF boost from the Breit-frame to the rest-frame  of the initial (i) and final (f)\\ nucleon states are given by:
\begin{equation}
  \label{Eq:defs0}
  \begin{aligned}
  &  \vec{k}^\text{i}_{i\perp} = \vec{k}_{i\perp} + \frac{\vec{q}_{\perp}}{2}x_i, \quad \vec{k}^\text{f}_{i\perp} = \vec{k}_{i\perp} - \frac{\vec{q}_{\perp}}{2}x_i; \quad i\neq j \,  \\ &\quad\text{and}\quad\vec{k}^\text{f(i)}_{j\perp} = (-) \frac{\vec{q}_\perp}{2}(1-x_j) -\sum_{i\neq j} \vec{k}_{i\perp}\, ,
  \end{aligned}
\end{equation} 
with the transverse momentum of the quark that absorbed the virtual photon being:
$  \vec k_{j\perp}=\pm\frac{\vec q_\perp}{2}-\sum^n_{i\neq j}\vec k_{i\perp}\,,$
with $+$ and $-$ meaning the  momentum in the initial and final hadron states, respectively.  For each Fock component of the LF wave function the transverse momenta add up to $\sum_{j=1}^n\vec k^\text{i}_{j\perp}=\sum_{j=1}^n\vec k^\text{f}_{j\perp}=\vec 0_\perp$, for the rest-frames of the initial and final hadron states. The normalized  proton wave function  gives $F(0) = 1$.

Beyond the electromagnetic processes, proton-\\proton collisions performed at the Large Hadron Collider in its high-luminosity phase  
requires  a detailed consideration of 
the nucleon structure
for the understanding of the observed data, associated with multiple parton interactions (MPIs), which are required for the description of hadronic final states (see e.g. \cite{Treleani:2017zzl}).

MPIs become more important for high-energy collisions as the  parton flux increases, while the parton momentum fractions decrease, as the nucleon momentum is shared among more participants. Therefore, the search for new physics demands the
consideration of MPIs in the dedicated experimental analysis (see the review book  on Multiple Parton Interactions at the LHC~\cite{Bartalini:2018qje}).
One example, is the double parton scattering (DPS) in hadron-hadron collisions, where two independent hard-scattering processes happen between partons from a  parton pair  in each hadron. It receives contributions from all LF Fock-components of each hadron wave function, and such information is encoded in the double parton distribution function~\cite{Blok_2011}.


The DPS cross section  which depends on the double parton distribution functions (dPDFs) contain contributions from all Fock-component of the wave function, and it is written as~\cite{Blok_2011}:
\newpage
\begin{equation}
\label{blokpdpdf}
\begin{aligned}
    D(&x_1,x_2,\vec \eta_\perp)= \sum_{n=3}^\infty
     D_n(x_1,x_2,\vec \eta_\perp)
 =\\&\sum_{n=3}^\infty   
   \int\frac{d^2k_{1\perp}}{(2\pi)^2}\frac{d^2k_{2\perp}}{(2\pi)^2}\,
    \Bigg\{ \prod_{i\neq 1,2} \int\frac{d^2k_{i\perp}}{(2\pi)^2} \int^1_0 dx_i\Bigg\}\, \\
    &\times\delta\left(1-\sum_{i=1}^n x_i\right)\, \delta\left(\sum_{i=1}^n\vec k_{i\perp}\right) \\
    &\times \Psi_n^\dagger(x_1,\vec k_{1\perp}+\vec \eta_\perp,x_2,\vec k_{2\perp}-\vec \eta_\perp,...)
    \\&\times\Psi_n(x_1,\vec k_{1\perp},x_2,\vec k_{2\perp},...)\, ,
\end{aligned}
\end{equation}
where  $\vec\eta_\perp$ is the transverse momentum shift and  for simplicity we have not depicted the polarization states for neither the constituents nor the hadron itself. 
The Fourier transform of $D(x_1,x_2,\vec \eta_\perp)$ in $\vec \eta_\perp$ gives the probability of finding the constituents 1 and 2 with momentum fraction $x_1$ and $x_2$ at a relative distance in the transverse direction $\vec y_\perp$ within the hadron state.  Here $\vec{y}_\perp$ is the Fourier conjugate of $\vec{\eta}_\perp$.
The quantity $ D_n(x_1,x_2,\vec \eta_\perp)$ is contribution from a given Fock-component of the LF wave function to the double parton distribution.

In terms of operator product the double distribution reads~\cite{Bali:2021gel}:
\begin{equation} 
\begin{aligned}
 &   \tilde D(x_1,x_2,\vec y_\perp)=2p^+\int dy^-
    \int\frac{dz^-_1}{2\pi}\frac{dz^-_2}{2\pi}\\&\times \text{e}^{i(x_1z^-_1+x_2z^-_2)p^+}
    \sum_{\lambda}\langle p,\lambda|\mathcal{O}(y,z_1)\mathcal{O}(0,z_2)|p,\lambda\rangle\, , 
    \end{aligned}
    \end{equation}
which has been obtained for the nucleon by recent  LQCD calculations for different operator structures~\cite{Bali:2021gel}. Despite such efforts, it is useful to obtain the dPDFs at the nucleon scale
 and
identifying properties of  the LF wave function, as for example using AdS/QCD approach~\cite{Traini:2016jru} and LF constituent quark  models (see e.g.~\cite{Rinaldi_2019}). Among such efforts to detail the LF wave function within a dynamical framework, we should mention the Basis Light-Front Quantization applied to QCD~\cite{Vary:2009gt} and recently used to study the nucleon~\cite{Mondal2020}.

Motivated by the above discussion, our goal in this work is to explore the  consequences of the relativistic LF three-body dynamics in the structure of the valence state of the nucleon, by studying one- and two-quark momentum distributions, where  effectively the interaction is dominated by a strong scalar quark-quark correlation. This model relies on the use of the contact interaction between the constituents within the Faddeev Bethe-Salpeter approach on the light-front~\cite{Frederico92,deAraujo95,Suisso_2002,Carbonell03}, and more recently the model was extended beyond the valence state in Euclidean space~\cite{Ydrefors:2017nnc} and in Minkowski space~\cite{Ydrefors:2019jvu,Ydrefors:2020duk}. For practical applications, the profile of the IR momentum dependence of the valence wave function, as for instance computed from the transverse amplitude,
obtained  by both the LF truncation and the full four-dimensional approach are essentially equivalent, once the bound state binding energies are close~\cite{Ydrefors:2017nnc}, which will be enough for the present  study. As a note, we observe that
short-range correlations between two quarks are present in the model, in analogy with the nucleon-nucleon short-range correlations~(see e.g. \cite{Hen_2017}), which have also its counterpart in the relativistic three-body wave function~\cite{Ydrefors:2020duk}, with the proviso that the UV behavior has to be viewed with caution as the scaling laws~\cite{Ji:2003fw} from QCD are not built-in.

The model
emphasizes the IR dynamics of constituent quarks with a dominant scalar diquark correlation.
 Indeed, one main feature that the continuum approaches to QCD have been teaching us is that two-quark correlations, namely, diquarks, which are not asymptotic states, are known to play a relevant role in the structure and dynamics of the nucleon (see e.g. \cite{Eichmann:2016yit} and the recent review~\cite{BarabanoPPNP2021}). 
 
 We should remind that the successful Nambu-Jona-Lasinio model applied to investigate phenomenological aspects of QCD in the IR region~\cite{Klevansky:1992qe}, embodies the dynamical chiral symmetry breaking by producing massive constituent ($m\sim 300$ MeV) for the $u$ and $d$ quarks and pions/kaons as Goldstone bosons, bringing in addition diquarks, with the favored one being the scalar color antitriplet $([ud]^{\overline 3_c}_{0^+})$ state. We see the renewed 
 interest from LQCD groups in determining the properties of diquarks in a gauge invariant way~\cite{Francis:2021vrr} gives  at the physical pion mass a difference of  319(1)~MeV between the mass of the lightest diquark, $[ud]^{\overline 3_c}_{0^+}$, and the light antiquark, and a size of about one fm. The low-energy diquark effective degree of freedom has also been invoked to smooth the transition between the hadron to quark phases of dense matter (see Ref.~\cite{Fukushima:2021ctq}).

 The model adopted in this work considers 
  a bound  or a virtual state pole in the quark-quark transition amplitude  as the main dynamical characteristic, which is in line with modern evidences of the relevance of the $[ud]^{\overline 3_c}_{0^+}$ state  in the IR properties of the  quark-quark effective interaction within the nucleon. We aim to explore the proton bound-state structure in terms of constituent quarks degrees of freedom by calculating the valence LF wave function, where our focus is to study its Ioffe time representation, as well as the different one- and two-quark momentum distributions.

 The rest of this work is organized as follows. A brief presentation of the LF three-quark model is given in section \ref{sec2}, containing 
the description of the homogeneous LF Faddeev integral equation and numerical results for the vertex Faddeev component. The 
results for the distribution amplitude and Ioffe-time image of the proton are given in section \ref{sec3}.   The calculations of the valence Dirac form factor of the proton are discussed in \ref{sec4}. The results for the momentum distributions, namely
valence parton distribution, valence double parton distribution and transverse momentum distributions for single and two quarks  in the forward limit and integrated in longitudinal momenta
are shown in section \ref{sec5}. The main points of our work are summarized in section \ref{secsummary}.  The work is completed by two appendices: in Appendix~\ref{App:Derivation} it is given the  derivation of the main dynamical integral equation of the model and in Appendix~\ref{App:NM} the adopted numerical method to solve it.

\section{Brief presentation of the LF three-quark model} \label{sec2}

 The  effective LF three-body model~\cite{Frederico92} that will be applied in the present work to study the proton, was originated as an attempt to generalize Weinberg's infinite momentum frame realization of the two-boson  BS equation~\cite{Weinberg:1966jm} to the three-body problem. Weinberg's original proposal kept a close relation of the three-dimensional dynamics in the light-front to the  Minkowski space BS equation for the bound state. Later on, within the framework of LF quantization  such an equation  was realized to be the lowest-order equation with the kernel expressing  the coupling of the two- and three-particle LF Fock-states (see e.g.~\cite{LepPRD1980}). The full equivalence of the two-body BS equation  and its LF three-dimensional representation have to take into account, besides the valence component, an infinite number of Fock-states. In principle, covariance under kinematical boosts is guaranteed even working with a finite truncation of  the Fock-space, however covariance under dynamical LF boosts, which are non-diagonal in the LF Fock-space, requires the dynamics to involve an infinite set of these states. Therefore, the  BS equation has built-in dynamically an infinite set of LF Fock-states. 
One possibility of projecting the BS equation to LF was done using the quasi-potential approach~\cite{Sales:1999ec}, where all the dynamics is buried in an effective interaction which contains the  virtual propagation of the system in an  infinite number of Fock-states, in close relation to the "iterated resolvent method"~\cite{Brodsky:1997de} to reduce the QCD dynamics in a hadron to its  valence component. 

At the three-body level, relevant to study the nucleon structure,  the counterpart of the Weinberg's equation for the contact interaction was proposed in~\cite{Frederico92}. It was performed the   projection to the LF of the Minkowski space Faddeev-Bethe-Salpeter (FBS) equation for the three-boson vertex  keeping only the valence contribution. In principle, the  interaction kernel of the LF-FBS equation  contains contributions beyond the valence component, appearing  as effective LF three-body forces (see e.g.~\cite{Ydrefors:2017nnc}). Indeed, the projection of the Minkowski space FBS equation  onto the LF was done via the quasi-potential approach in~\cite{Marinho:2007zz} and further developed in~\cite{Frederico:2010zh,Guimaraes:2014kor}. We outline in  Appendix~\ref{App:Derivation} the derivations including the three-body valence wave function and LF dynamical equation  within the three-body BS framework.

 As we have mentioned above, the model adopted to investigate the Ioffe time representation of the wave function and also the double parton distribution, is based on the contact interaction between the constituent quarks, where the spin degree of freedom is not taken into account, as it is our aim to to study the spatial non-polarized distribution of the quarks in the valence state. In the model we consider only  the totally symmetric momentum part of the
the colorless three-quark wave function corresponding to the  valence nucleon state, as we are interested  for the time being on the investigation of the properties associated with the momentum distributions and the image of the nucleon onto the null-plane. 
The  valence LF wave function is given by~\cite{Ydrefors:2020duk}:
\begin{equation}
\label{Eq:BS_wf}
\begin{aligned}
 & \Psi_3(x_1,\vec{k}_{1\perp}, x_2, \vec{k}_{2\perp}, x3, \vec{k}_{3\perp}) = \\&\frac{ \Gamma(x_1, k_{1\perp}) + \Gamma(x_2, k_{2\perp}) + \Gamma(x_3, k_{3\perp}, )}{\sqrt{x_1 x_2 x_3}(M_N^2 - M^2_0(x_1,\vec{k}_{1\perp},x_2, \vec{k}_{2\perp},x_3, \vec{k}_{3\perp}))},
\end{aligned}
\end{equation} 
 with $\Gamma(x_i, k_{i\perp})$, where $k_{i\perp}=|\vec{k}_{i\perp}|$,   being the Faddeev component of the vertex function for the bound state, $x_1 +x_2 +x_3=1$, 
$\vec{k}_{1\perp}+\vec{k}_{2\perp}+\vec{k}_{3\perp}=\vec 0_\perp$ and
\begin{equation}
\begin{aligned}
  M^2_0(x_1&,\vec{k}_{1\perp},x_2, \vec{k}_{2\perp}, x_3,\vec{k}_{3\perp}) = \\&\frac{\vec{k}_{1\perp}^2 + m^2}{x_1} + \frac{\vec{k}_{2\perp}^2 + m^2}{x_2} + \frac{\vec{k}_{3\perp}^2 + m^2}{x_3},
 \end{aligned}
 \end{equation}
 is the free three-body squared mass for on-mass-shell constituents.  The factorized form of the valence wave function, namely with a vertex function depending solely on the bachelor quark LF momenta, is a consequence of the effective contact  interaction between the constituent quarks, which is an idealized model resembling the successful Nambu-Jona-Lasinio model applied to model QCD~\cite{Klevansky:1992qe}. It should be understood as an effective low-energy model which is meant to have significance in the IR region where constituent quarks are massive and bound forming the nucleon.  We show in  Appendix~\ref{App:Derivation} the derivation of the valence LF wave function starting from the three-legs Bethe-Salpeter amplitude. 

\subsection{Homogeneous LF Faddeev integral equation}
\label{sec2a}

The Faddeev equation for the vertex component of the valence LF wave function
 is given by \cite{Frederico92,Carbonell03}:
\begin{equation}
  \label{Eq:3b_LF}
  \begin{aligned}
  \Gamma(x, k_\perp) = & \frac{\mathcal{F}(M^2_{12})}{(2\pi)^3}\int_0^{1-x}\frac{dx'}{x'(1-x-x')}\\ & \times
  \int_0^\infty d^2 k'_\perp 
  \frac{\Gamma(x',k'_\perp)}{\widehat M_0^ 2 - M_N^2},
\end{aligned}
\end{equation}
where 
\begin{equation}
\widehat M_0^ 2=M^2_0(x,\vec{k}_\perp,x',\vec{k}'_\perp, 1-x-x', -(\vec{k}_\perp+\vec{k}'_\perp))\, .
\end{equation}
 For the sake of completeness,  we provide in  Appendix~\ref{App:Derivation} a derivation of Eq.~\eqref{Eq:3b_LF} starting from the three-boson BS equation by  projecting it onto the LF via the quasi-potential technique.

The two-quark amplitude has the expression~\cite{Ydrefors:2020duk}:
 \begin{multline}
\label{Eq:F_amp}
\mathcal{F}(M^2_{12})=
\frac{\Theta(-M_{12}^2)}{\frac{1}{16\pi^2 y}\log\frac{1+y}{1-y}-\frac{1}{16\pi m a}}
\\ 
+\frac{\Theta(M_{12}^2)\,\,\Theta(4m^ 2-M_{12}^2)}{\frac{1}{8\pi^2 y'}\arctan y'-\frac{1}{16\pi m a}}\,,
\end{multline}
with its argument, the effective off-shell mass of the two-quark subsystem squared, given by
\begin{equation}
\begin{aligned}
&M^2_{12} = (1-x)M^2_N - \frac{k^2_{\perp}+ (1-x)m^2}{x}\, , \\
& y=\frac{M_{12}}{\sqrt{4m^2 - M^2_{12}}}\quad\text{and}\quad y'=\frac{\sqrt{-M^2_{12}}}{\sqrt{4m^2 - M^2_{12}}}\, . 
\end{aligned}
\end{equation}
 Additionally, in Eq.~\eqref{Eq:F_amp}, $\Theta(x)$ denotes the Heaviside theta function.

The kernel of the LF Faddeev equation~\eqref{Eq:3b_LF} contains the quark exchange mechanism  expressed by the presence of  the three-quark LF resolvent, namely the operator $[\widehat M^ 2_0-M_N^2]^{-1}$.  Consistently with the adopted model, it is well known~\cite{Eichmann:2016yit} that the four-dimensional formulation of the three-quark BS equation presents the quark exchange kernel, when the diquarks dominates the quark-quark interaction.  We should emphasize the physical significance of the present model in the context of  nucleon models formulated commonly within  the BS approach in Euclidean space~\cite{Eichmann:2016yit}.  Our model provides directly the LF wave function allowing to access  momentum distributions and keeps the straight relation with the Bethe-Salpeter framework, in contrast with commonly used Euclidean BS approaches. Furthermore, it incorporates the main physics of more sophisticated Euclidean formulations, as the quark exchange kernel, and a pole in the quark-quark amplitude representing a bound or virtual diquark state.

The quark-quark scattering amplitude, $\mathcal{F}(M^2_{12})$, weights the Faddeev LF integral equation and carries the pole of the  bound or virtual diquark, that depends on the scattering length $a$, which can be either positive or negative.
If $a>0$, the quark-quark system is bound and the nucleon will be described as a quark-diquark system. On the contrary if $a$ is negative no physical two-body bound-state exists and the nucleon is thus a Borromean state. In both cases, $\mathcal{F}(M^2_{12})$ has a pole, for $a>0$  in the physical complex-energy sheet and for $a<0$ in the $2^\text{nd}$ sheet, meaning the virtual state. Therefore, in either case the strong diquark correlation is present in the model and should be interpreted as dominating the IR properties of the nucleon. Both of these two cases will be investigated in this paper. An earlier study of the nucleon performed with a truncated form of Eq.~\eqref{Eq:3b_LF} was performed in~\cite{deAraujo95}.

 In Appendix~\ref{App:NM}  it is explained the adopted numerical method to solve the integral equation  by using a bicubic spline expansion. The condition $F(0)=1$, has been adopted to normalize the solution, where $F(Q^2)$ is the valence Dirac form factor, which will be discussed in Sec.~\ref{sec4}.
\begin{table}
\centering
\begin{tabular}{c c c c c c}
    \toprule
    Model & $m$  & $a$  & $M_2$  & $M_N$ &  $r_{F_1}$
     \\
&    [MeV] &  $[m^{-1}]$ & [MeV] & [$m$] & [fm]\\
    \midrule
     I &  317 & -1.84 & - & 2.97 &  0.97\\
    II &  362  & 3.60 & 681 & 2.60 & 0.72 \\
     \bottomrule
\end{tabular}
\caption{Constituent mass in MeV, scattering length, diquark mass, and three-body mass for the two considered models. Our diquark mass for model II is slightly smaller than the scalar diquark mass of 691 MeV obtained in \cite{Ferretti19}. Also shown is the radius defined as $r_{F_1} = \hbar c \sqrt{-6\frac{dF_1}{dQ^2}|_{Q^2=0}}$ and the corresponding experimental value  is $0.757$ fm \cite{Xiong19}. \label{Tab:Table_1}}
\end{table}

\subsection{Vertex Faddeev component} 
\label{sec2b}

The structure of the three-quark valence state is encoded in the vertex function $\Gamma(x, k_\perp)$, which was computed with the two parameter sets from Table~\ref{Tab:Table_1}. The constituent quark masses are  317 MeV (model I) and 362 MeV (model II) to be compared with about 350 MeV from a recent LQCD calculation in the Landau gauge~\cite{Oliveira:2018lln}. 
We choose two possibilities for diquarks, namely an unbound one for $a<0$ and a bound one for $a>0$, with a diquark mass of 681 MeV.
These parameters are found by reproducing qualitatively the space-like Dirac form factor up to about 1 GeV$^2$, as it will be shown later on. We observe that, the diquark  mass of 681 MeV, which has a difference of 319~MeV with respect to the quark mass,  coincidentally  matches the gauge invariant result from the LQCD calculation~\cite{Francis:2021vrr} of 319(1)~MeV at the physical pion mass.

\begin{figure}[bht]
  \centering
  \includegraphics[scale=0.86]{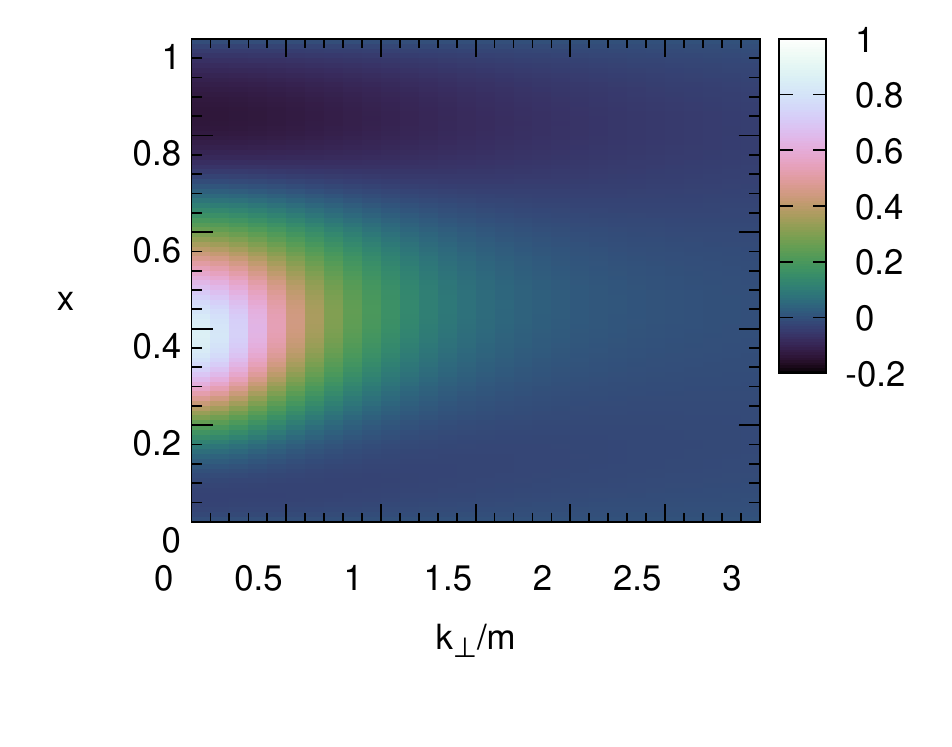}
  \includegraphics[scale=0.86]{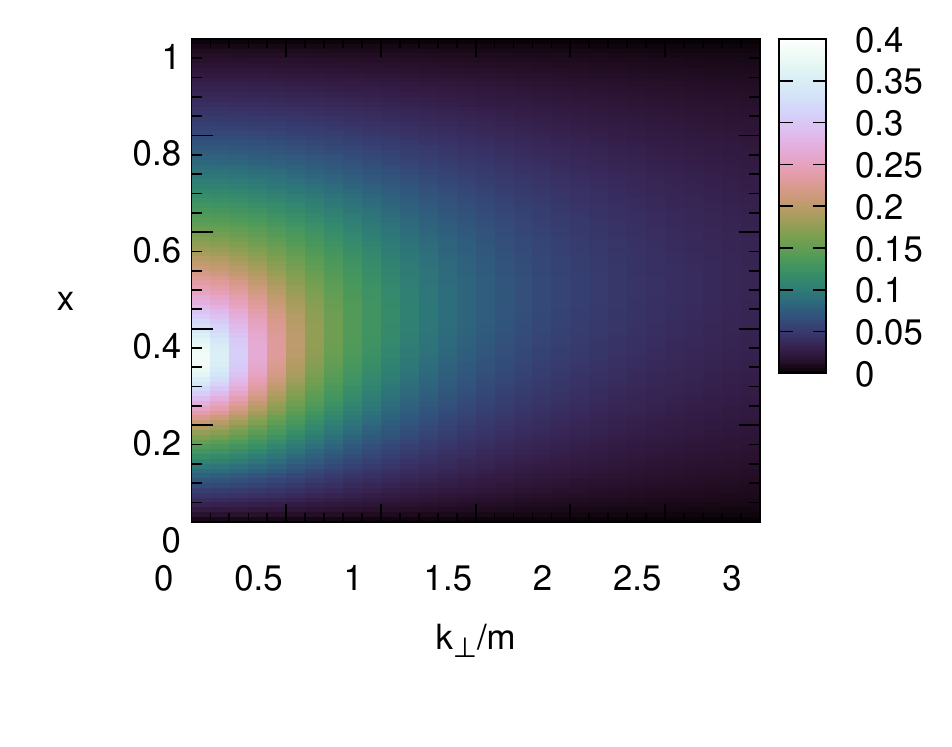}
  \caption{Vertex function, $\Gamma(x,k_\perp)$, for the models I (lower panel) and II (upper panel). \label{Fig:Gamma}}
\end{figure}

The results for the vertex function $\Gamma(x, k_\perp)$ are shown in Fig.~\ref{Fig:Gamma} for models I (lower panel) and II (upper panel).
The vertex function for both models has characteristic transverse momentum around the IR scale of $\sim\Lambda_{QCD}$, which drives the decreasing behaviour with $k_\perp$. 
In addition, the vertex function peaks between $x\sim 0.35-0.4$, and the peak evolves to somewhat larger values of $x$ with $k_\perp$, as a consequence of the dominance of $k^2_\perp/x$ in the free squared mass operator, which comes with the quark exchange kernel, as should be a general feature of the diquark (bound or virtual) dominance  in the quark-quark interaction.

It is seen in the  upper panel of Fig.~\ref{Fig:Gamma} that for model II with $a>0$, there is one node around $x=0.8$ and also one for small $x$. As studied in detail in Ref.~\cite{Ydrefors:2017nnc},
  for $a>0$ the physical ground state of  \eqref{Eq:3b_LF} is not the lowest energy solution of the equation. That is, it exists another unphysical solution with $M^2_N < 0$.
This state is the relativistic analog of the well-known Thomas collapse in non-relativistic three-body systems with zero-range interaction~\cite{ThomasPR1935}. For example, at $1/(a\,m) = 0.26$ one has $M_N^2 = - 69 \: m^2$, so it is a very deep state. In principle,  it should be possible to remove this state by  a momentum cut-off $\Lambda$ of the order of 1~GeV, which would weaken the interaction in the short-range region.

\section{Distribution amplitude and Ioffe-time image}
\label{sec3}

\subsection{Distribution amplitude} 
\label{sec3a}

\begin{figure}[thb]
\centering
\includegraphics[scale=0.83]{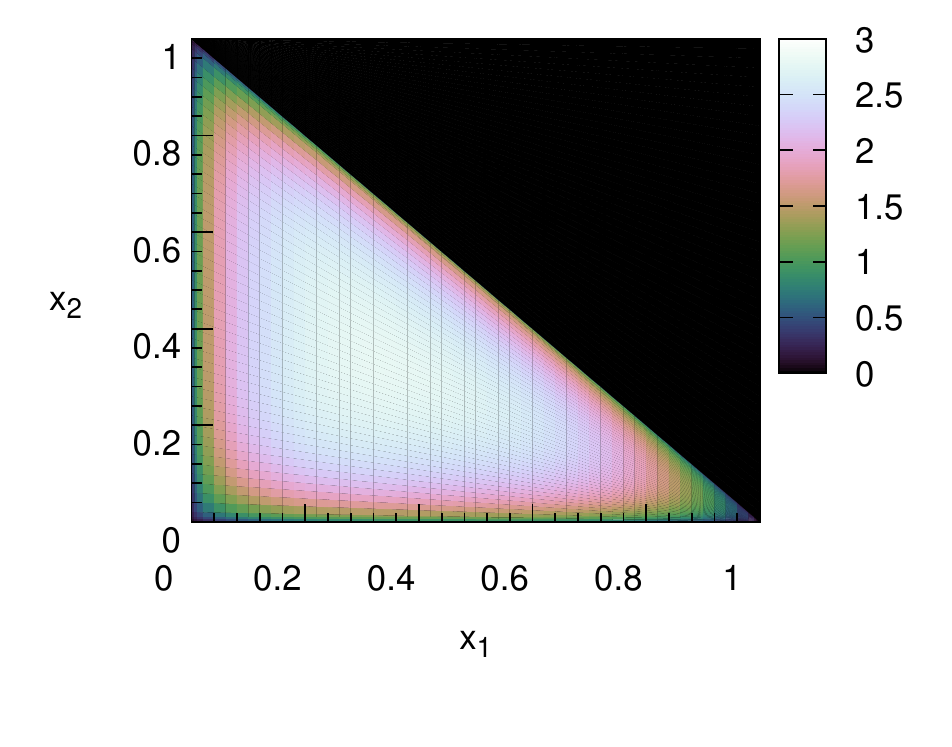}
\includegraphics[scale=0.83]{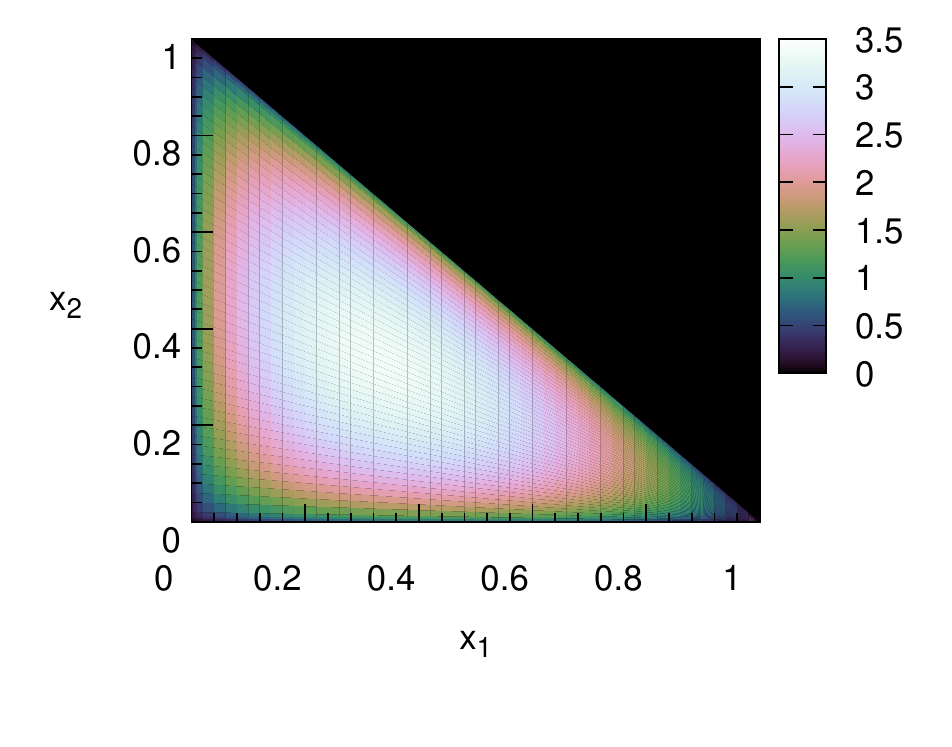}
\vspace{-1cm}
\caption{Distribution amplitude as a function of $x_1$ and $x_2$, for the model I (lower panel) and for model II (upper panel). \label{Fig:da}}
\end{figure}

The distribution amplitude (DA), $\phi(x_1,x_2)$ for the nucleon is defined as
\begin{equation}
\begin{aligned}
  \phi(&x_1,x_2,x_3)= \int d^2 k_{1\perp} d^2 k_{2\perp}d^2 k_{3\perp}\\
  &\times \delta(\vec{k}_{1\perp} + \vec{k}_{2\perp} + \vec{k}_{3\perp}) \Psi_3(x_1, \vec{k}_{1\perp},x_2, \vec{k}_{2\perp}, x_3, \vec{k}_{3\perp}),
  \end{aligned}
\end{equation} with $x_3 = 1 - x_1 - x_2$,  and obeys the  symmetry relation
\begin{equation}
    \phi(x_1,x_2,x_3)=\phi(x_2,x_1,x_3).
    \end{equation}
 It gives the dependence of the wave function on the longitudinal momentum fraction when  the quarks share the same transverse position.  %
In Fig.~\ref{Fig:da}, the calculated DA is shown for the two models considered in the present work. In the figure the DA was normalized so that 
\begin{equation}\begin{aligned}&\int_0^1 dx_1 \int_0^{1-x_1}dx_2 \int_{0}^1 dx_3 \delta(1 - x_1 - x_2 - x_3)\\ &\times \phi(x_1,x_2,x_3)
 = 1 \nonumber.
\end{aligned}\end{equation} 
The two different models give similar results with a slightly wider distribution for model II, which reflects the wave function behaviour close to $x_i\sim 0$.  Further insight comes with the Fourier transform as  discussed in what follows.

 \subsection{  Ioffe-time image of the valence state} 
\label{sec3b}

 For the study of the space-time structure of the proton it is of interest to obtain the wave function in terms of the Ioffe-times ($\tilde{x}_1$ and $\tilde{x}_2$) and the transverse coordinates ($\vec{b}_{1 \perp}$ and $\vec{b}_{2\perp}$), which provides the image of the proton on the null-plane $x^+=0$.   Such study has been performed recently for the pion~\cite{dePaula:2020qna}. This is accomplished through the Fourier transform of $\Psi_3(x_1,\vec{k}_{1\perp};x_2,\vec{k}_{2\perp};x_3,\vec{k}_{3\perp})$. For simplicity, we consider here the particular case:
\begin{equation}
  \begin{aligned}
   \Phi(\tilde{x}_1&, \tilde{x}_2)\equiv\tilde{\Psi}_3(\tilde{x}_1,\vec 0_\perp, \tilde{x}_2,\vec 0_\perp)= \\
   &\int_0^1 dx_1 \, \text{e}^{i\tilde{x}_1\, x_1} \int_0^{1 - x_1} dx_2\int_0^1 dx_3 \\
   & \times \delta(1 - x_1 - x_2 - x_3)\, \text{e}^{i\tilde{x}_2\, x_2}\,\phi(x_1, x_2 ,x_3)\,,
     \label{Eq:ioffe}
  \end{aligned}
\end{equation}
 where the configuration space wave function is computed at the origin $\vec{b}_{1\perp} = \vec{b}_{2\perp} = \vec 0_\perp$.

\begin{figure}[thb]
\centering
\includegraphics[scale=.83]{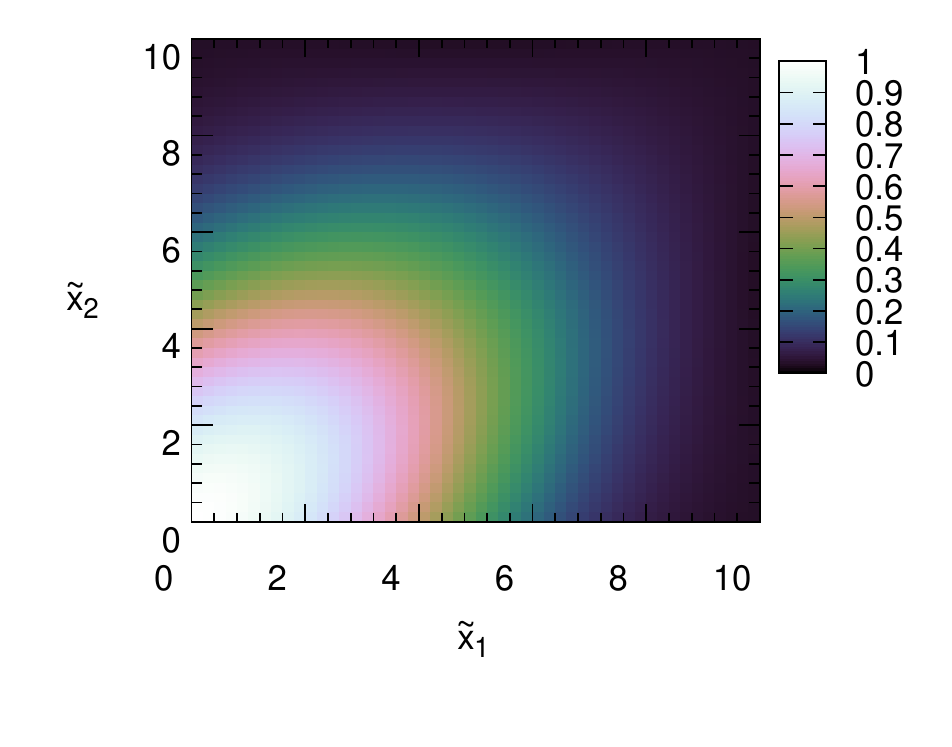}
\includegraphics[scale=0.5]{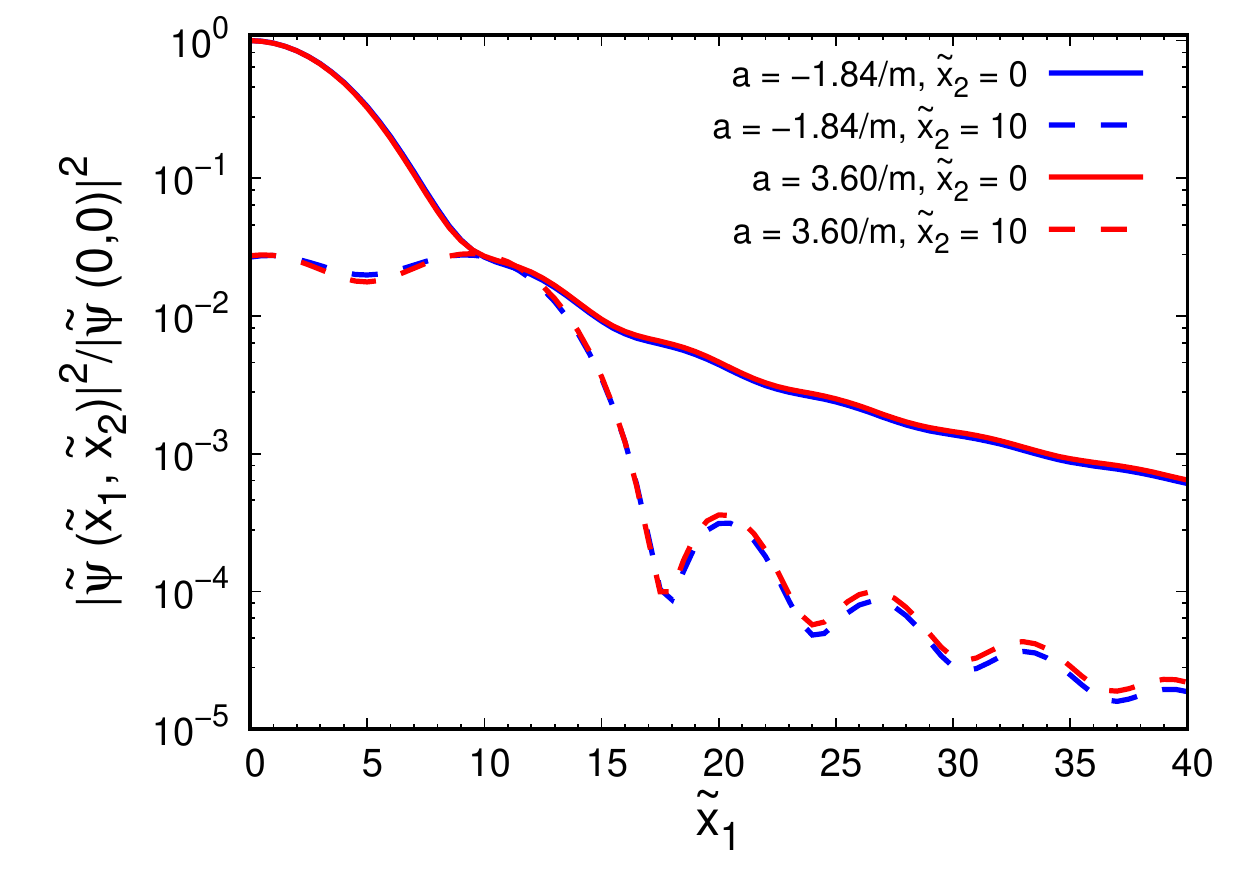}
\caption{Upper panel: Squared modulus of the Ioffe-time distribution as a function of $\tilde{x}_1$ and $\tilde{x}_2$, for the model I.
Lower panel: Squared modulus of the Ioffe-time distribution as a function of $\tilde{x}_1$ for two fixed values of $\tilde{x}_2$, namely $\tilde{x}_2$ = 0 (solid line) and $\tilde{x}_2$ = 10 (dashed line). Results shown  for the model I (blue line) and model II (red line).\label{Fig:ioffe}}
\end{figure}

 In Fig.~\ref{Fig:ioffe} we present our results for the squared modulus of the Ioffe-time distribution given by Eq.~\eqref{Eq:ioffe}. In the upper panel is shown the 3D plot of the distribution in terms of the variables $\tilde{x}_1$ and $\tilde{x}_2$ for model I. It is clear the preference of quarks to minimize the relative distance in Ioffe time, as also observed along  $\tilde x_1=\tilde x_2$. The decrease along the just reflects the presence of the third quark that recoils as the center of mass is at rest. Notably, there is no perceptible difference in this plot between model I and II. 
 
 Then, in the lower panel of Fig.~\ref{Fig:ioffe} we show for both models the Ioffe-time distribution as a function of $\tilde{x}_1$ for two fixed values of $\tilde{x}_2$, namely $\tilde{x}_2=0$ and $\tilde{x}_2=10$.  It is seen that the results obtained with the two parameter sets are almost identical for $\tilde{x}_1<17$. In addition, we observe the equality between the $|\tilde\phi(0,10)|=|\tilde\phi(10,0)|=|\tilde\phi(10,10)|$, which comes from the permutation symmetry of the wave function:
 \begin{equation}
   |\tilde\phi(\tilde x,0)|=|\tilde\phi(0,\tilde x)|=|\tilde\phi(\tilde x,\tilde x)|\, , 
 \end{equation}
 which is a general characteristic of the model. This explains also the rough flat behavior of $|\tilde\phi(\tilde x_1,10)|$ for $0<\tilde x_1< 10$ (dashed line).

 For $\tilde{x}_2 = 10$ (dashed lines in Fig.~\ref{Fig:ioffe}) a sizable decrease of the magnitude is observed at $\tilde{x}_1 > 10$ and for larger values an oscillatory behavior is seen. This reflects the size of the proton of about 1~fm  and a mass of 1~GeV, with their dimensionless product taking into account the factor 1/2 from the adopted metric $\lambda^{-1}\sim 0.1$ with the characteristic oscillatory pattern having a wave length in Ioffe time of about 10, which is observed in the figure. This is also roughly the dimensionless scale, which governs the decrease of the wave function in the two quarks relative separation in Ioffe time.

 Both trends, namely oscillation and damping of the wave function  are essentially the same for models I and II, as their proton charge radius are somewhat close, besides the same proton mass, as seen in Table~\ref{Tab:Table_1}. It suggests that this general behavior should be  quite model independent. Notably, a similar qualitative behavior of the Ioffe-time distribution for the pion was obtained in \cite{dePaula:2020qna}. We observe an exponential damping of the probability density with the relative separation between the Ioffe time of the two quarks, and the damping is expected to be more sizable if confinement is incorporated as it is effective at large  distances.

\section{ Valence Dirac form factor of the proton}
\label{sec4}

In the three-body null-plane model, i.e.~only taking into account the valence contribution $(n=3)$ in the form factor formula~\eqref{Eq:G_EFS}, the Dirac  form factor is given by:
\begin{small}
\begin{equation}
  \label{Eq:G_E}
  \begin{aligned}
  F_{\text{1}}&(Q^2) = \Bigg\{ \prod_{i=1 }^3\int\frac{d^2k_{i\perp}}{(2\pi)^2} \int^1_0 dx_i\Bigg\}\, \delta\left(1-\sum_{i=1}^3 x_i\right)\, \\ &\times\delta\left(\sum_{i=1}^3\vec k^\text{f}_{i\perp}\right)
  \Psi_3^\dagger(x_1,\vec{k}^\text{f}_{1\perp}, ...)\Psi_3(x_1,\vec{k}^\text{i}_{1\perp},...),
\end{aligned}
\end{equation}
\end{small}
with $Q^2 = \vec{q}_\perp \cdot \vec{q}_\perp$. 
In Eq.~\eqref{Eq:defs0} the transverse momentum of the quarks in the Breit frame are given.

One has in Eq.~\eqref{Eq:G_E} that $d^2k_{i\perp} = |\vec{k}_{i\perp}|d|\vec{k}_{i\perp}|d\theta_{i}$ ($i = 1, 2$) with  $\vec{k}_{i\perp}\cdot \vec{q}_{\perp} = |\vec{k}_{i\perp}||\vec{q}_\perp|\cos \theta_i$.
Additionally, the needed magnitudes of the transverse momenta are given by
\begin{equation}
 \begin{aligned}
 \Bigl|\vec k^\text{f(i)}_{i\perp}\Bigr|^2&=\Bigl|\vec{k}_{i\perp} \pm \frac{\vec{q}_\perp}{2}x_i\Bigr|^2 
 \\ & =\vec{k}_{i\perp}^{2} + \frac{Q^2}{4}x_i^2 \pm x_i|\vec{k}_{i\perp}||\vec{q}_\perp|\cos\theta_i
 \, ,
\end{aligned}
\end{equation}
with $-$ for $\text{f}$ and $+$ for $\text{i}$, in addition we have
\begin{small}
\begin{equation}
  \label{Eq:mag_k3_if}
  \begin{aligned}
\Bigl|&\vec k^\text{f(i)}_{3\perp}\Bigr|^2=   \Bigl|\pm \frac{\vec{q}_\perp}{2}(x_3-1) - \vec{k}_{1\perp} - \vec{k}_{2\perp}  \Bigr|^2 = 
\\    & (1-x_3)^ 2\frac{Q^2}{4} + \vec{k}_{1\perp}^{2} +  2|\vec{k}_{1\perp}||\vec{k}_{2\perp}|\cos(\theta_1 - \theta_2)
    \\    &  + \vec{k}_{2\perp}^{2}  \pm (1-x_3)|\vec{q}_\perp| \bigl(|\vec{k}_{1\perp}|\cos\theta_1 + |\vec{k}_{2\perp}|\cos\theta_2\bigr)\, .
  \end{aligned}
\end{equation}
\end{small}

\begin{figure}[thb]
  \centering
\includegraphics[scale=0.6]{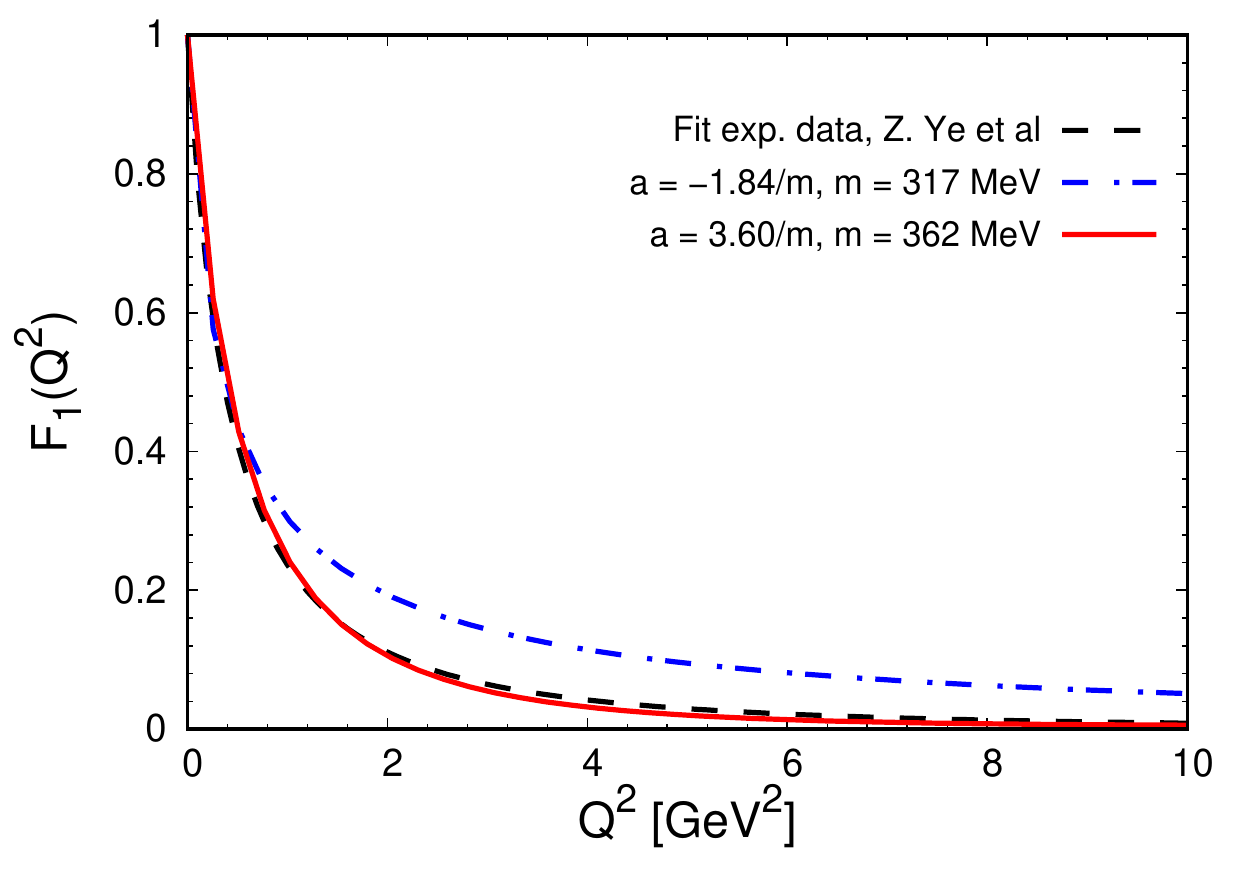} 
  \caption{
  Computed $F_1(Q^2)$ (solid and dot-dashed lines) compared with the empirical fit (dashed line) obtained in Ref.~\cite{Ye18}. 
  \label{Fig:FF}}
\end{figure}

In Fig.~\ref{Fig:FF}, the computed Dirac form factor, $F_1(Q^2)$, for the two parameter sets listed in Table~\ref{Tab:Table_1}, is compared with the global fit to experimental data by Ye et al \cite{Ye18}. It is seen that model II with $a = 3.60/m$ gives a quite good agreement with the experimental data. This thus favors the description of the nucleon as a quark-diquark system. The computed values of the radius $r_{F_1} = \hbar c \sqrt{-6\frac{dF_1}{dQ^2}|_{Q^2=0}}$ for the two considered models are also listed in Table \ref{Tab:Table_1}. The model II gives a radius of 0.72 fm which is about 5\% lower than the experimental value of  $0.757$ fm \cite{Xiong19} from the charge form factor. On the contrary, for the first model with $a < 0$  a rather large radius of 0.97 fm was obtained. We could have attempted to fit the charge radius from $F_1$, by changing the constituent quark mass and scattering length. However, we choose to keep the qualitative reproduction of the form factor up to  $Q^2\sim 1 $~GeV$^2$, which should be the scale of our model.

\section{ Momentum distributions}
\label{sec5}

\subsection{Valence parton distribution}
\label{sec5a}

We study next the decomposition of the single parton distribution function (PDF), obtained from the integrand of  Eq.~\eqref{Eq:G_E} of the Dirac form factor:
 \begin{equation}
  \label{Eq:pdf_Q20}
  \begin{aligned}
  {f}_1(x_1)& = \frac{1}{(2\pi)^6}\int_0^{1-x_1} d x_2  \int_0^1 dx_3 \delta(1 - x_1 - x_2 - x_3) 
  \\ & \times \int d^2 k_{1\perp} d^2 k_{2\perp}d^2 k_{3\perp}\delta(\vec{k}_{1\perp} + \vec{k}_{2\perp} + \vec{k}_{3\perp})\\
  &\times|\Psi_3(x_1,\vec{k}_{1\perp}, x_2, \vec{k}_{2\perp}, x_3, \vec{k}_{3\perp})|^2 
  \\ &= I_{11} + I_{22} + I_{33} + I_{12} + I_{13} + I_{23}.
\end{aligned}
\end{equation}
where the contributions to the PDF are defined for $i=1, 2, 3$ as:
\begin{equation}
  \begin{aligned}
    &I_{ii} = \frac{1}{(2\pi)^6}\int_0^ {1-x_1} dx_2 \int_0^1 \frac{dx_3}{x_1 x_2 x_3}\delta(1 - x_1 - x_2 - x_3)\\
    & \times \int d^2 k_{1\perp} d^2 k_{2\perp} d^2 k_{3\perp}\delta(\vec{k}_{1\perp} + \vec{k}_{2\perp} + \vec{k}_{3\perp}) \\
    & \times \frac{[\Gamma(x_i, \vec{k}_{i\perp})]^2}{(M_N^2 - M^2_0(x_1,\vec{k}_{1\perp},x_2,\vec{k}_{2\perp}, x_3, \vec{k}_{3\perp}))^2}\, , 
    \end{aligned}
  \end{equation}
  and for $i \neq j$:
\begin{equation}
  \begin{aligned}
   & I_{ij} = \frac{2}{(2\pi)^6}\int_0^{1-x_1} d x_2 \int_0^1 \frac{dx_3}{x_1 x_2 x_3}\delta(1 - x_1 - x_2 - x_3)\\
   & \times \int d^2 k_{1\perp} d^2 k_{2\perp}  d^2 k_{3\perp}\delta(\vec{k}_{1\perp} + \vec{k}_{2\perp} + \vec{k}_{3\perp}) \\
   & \times \frac{\Gamma(x_i, \vec{k}_{i\perp})\Gamma(x_j, \vec{k}_{j\perp})}{(M_N^2 - M^2_0(x_1,\vec{k}_{1\perp},x_2, \vec{k}_{2\perp}, x_3, \vec{k}_{3\perp}))^2} \, .
\end{aligned}
  \end{equation}
  Due to the symmetries of the three-body wave function under exchange of particles 2 and 3, it follows that $I_{22} = I_{33}$ and $I_{12} = I_{13}$.

\begin{figure}[thb]
  \centering
  \includegraphics[scale=0.5]{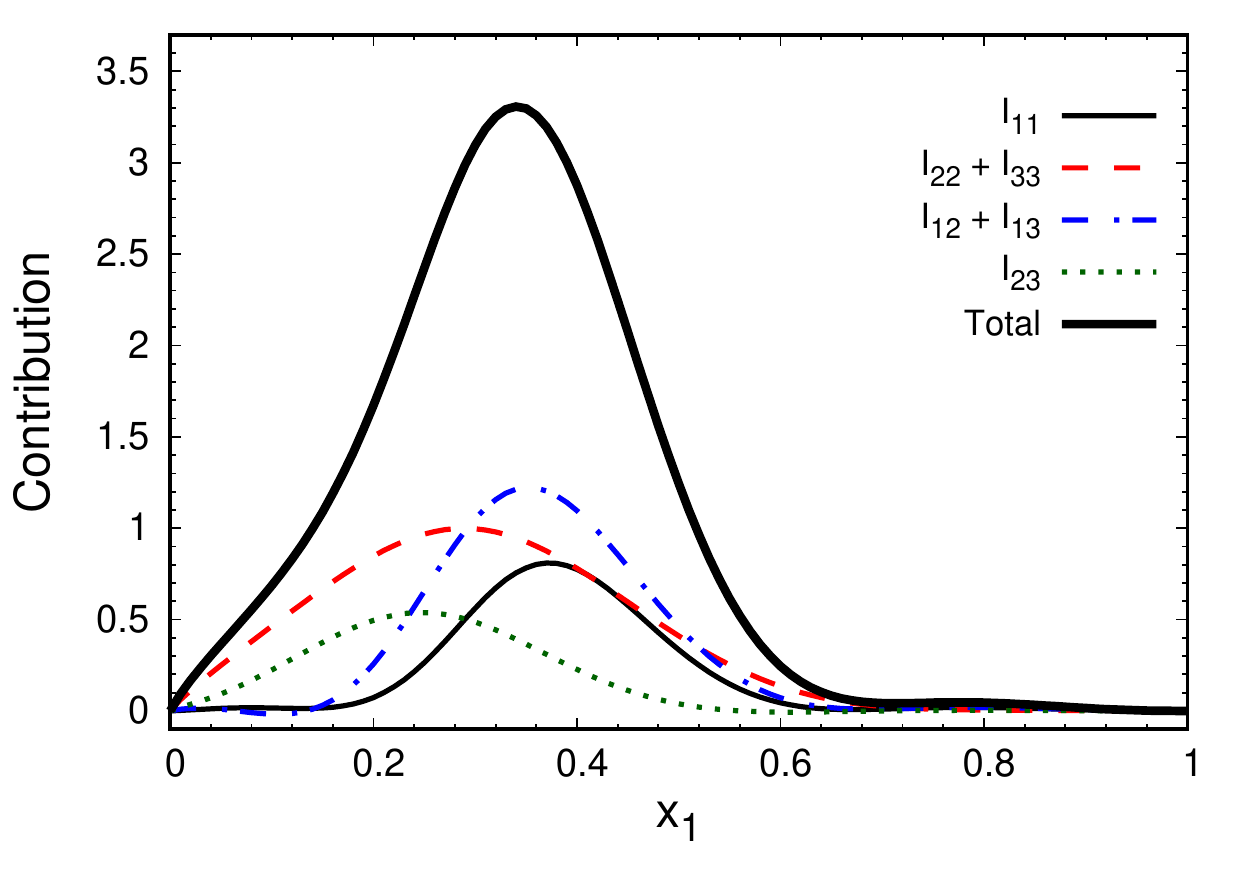}
 \includegraphics[scale=0.5]{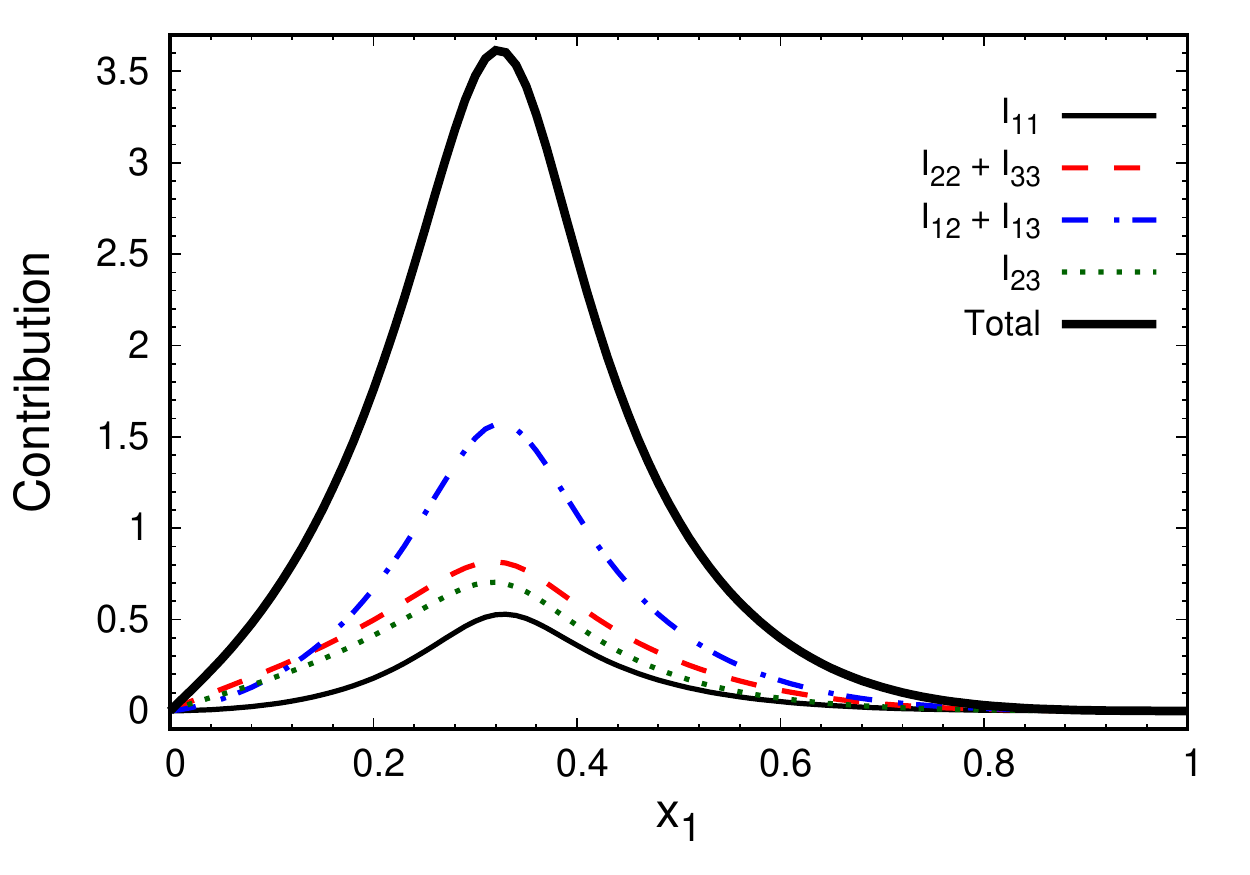}
  \caption{
  Contributions to the PDF  for the  models I (lower panel) and II (upper panel)
  \label{Fig:PDFs}}
\end{figure}

 The contributions to the PDF at vanishing $Q^2$  are presented for the two considered models in the middle and upper panels of Fig.~\ref{Fig:FF}. The total PDF is also shown in each panel with a thick solid line. For both models a maximum of the PDF is seen at $x\approx 1/3$. As is seen the right panel, the model II with  a positive scattering length gives an almost flat behavior  around $x=0.8$ for the PDF. Larger differences in the behavior of the contributions can also be observed for this set of parameters.  

Interesting to observe that all the contributions have about the same size, and peaks around 0.35, despite we are measuring the PDF for the quark labeled by 1 with momentum fraction $x_1$. More variation of the peaks position are seen for $a>0$ where the vertex function  has a node  for model II (see Fig.~\ref{Fig:Gamma}), while the interplay with the denominator of the wave function where the smallest virtuality in the mass squared leads to fixed positions in all contributions around $1/3$. The contribution from $I_{11}$ corresponding to a  configuration, where quark 1 is picked up while the pair of quarks interacts, does not dominate, meaning that the symmetrization of the momentum component of the wave function is crucial for the proton PDF.

\subsection{ Valence double parton distribution}  
\label{sec5b}

Following  Eq.~\eqref{blokpdpdf},
 we write the valence contribution to the double quark distribution function as:
  \begin{equation}
  \begin{aligned}
   & D_3(x_1, x_2; \vec \eta_\perp) = \frac{1}{(2\pi)^6}\int_0^1 dx_3 \delta(1 - x_1 - x_2 - x_3)\\
   &\times \int d^2 k_{1\perp} d^2 k_{2\perp} d^2 k_{3\perp}\delta(\vec{k}_{1\perp} + \vec{k}_{2\perp} + \vec{k}_{3\perp})\\
    &\times \Psi_3^\dagger(x_1,\vec{k}_{1\perp} + \vec \eta_\perp; x_2, \vec{k}_{2\perp} - \vec \eta_\perp; x_3, \vec{k}_{3\perp})\\
    &\times\Psi_3(x_1,\vec{k}_{1\perp}; x_2, \vec{k}_{2\perp}; x_3, \vec{k}_{3\perp}).
  \end{aligned}
  \end{equation}

\begin{figure}[thb]
  \centering
 \includegraphics[scale=0.84]{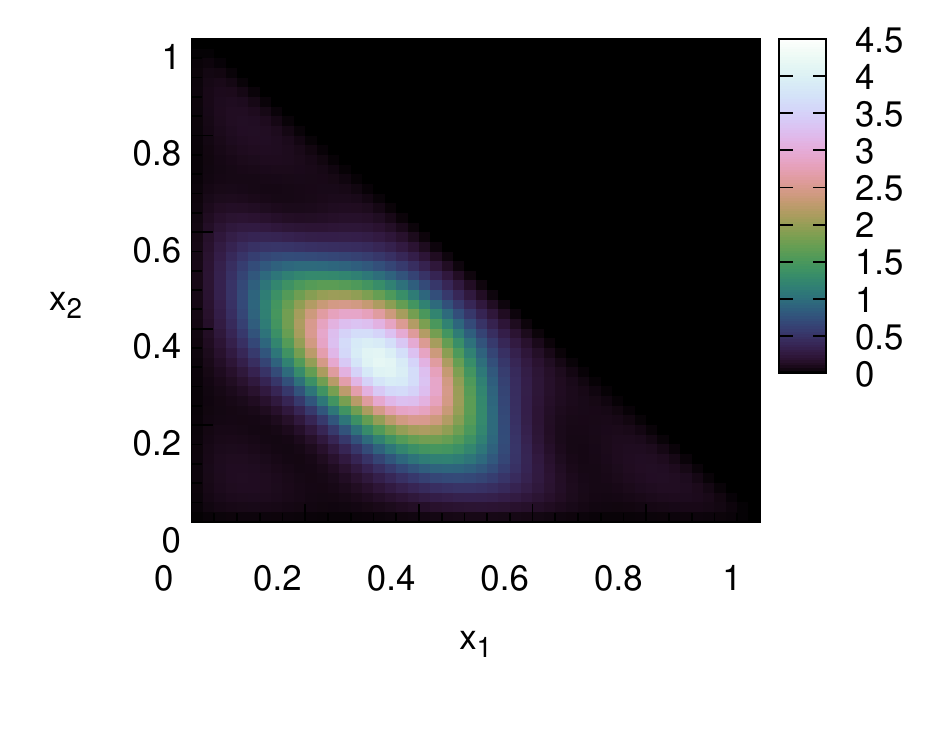}
 \includegraphics[scale=0.84]{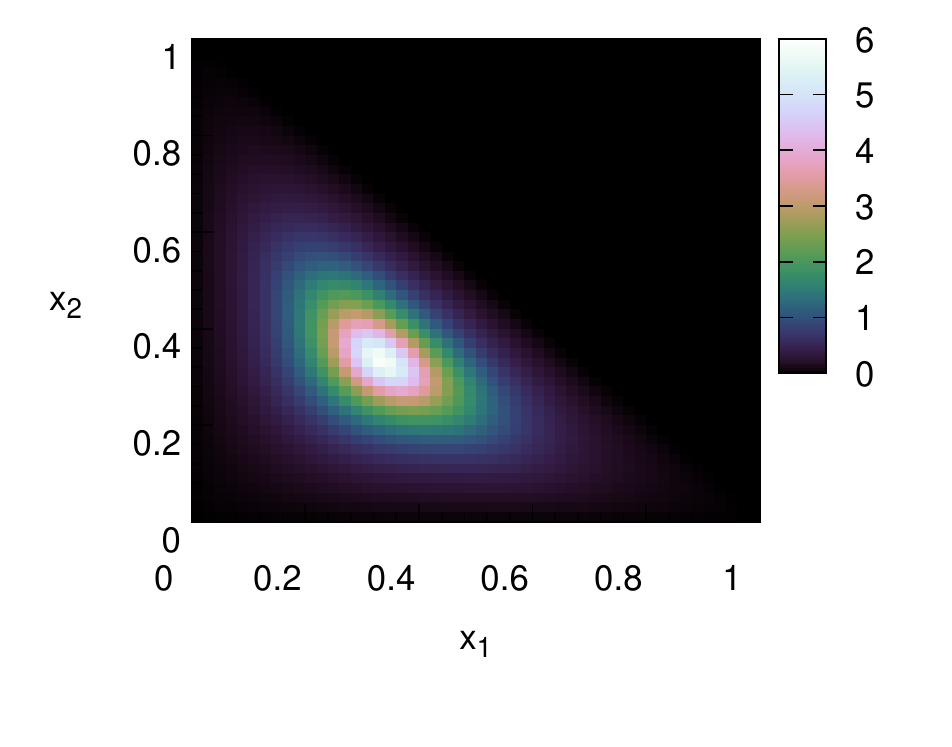}
 \vspace{-1cm}
  \caption{Double quark distribution function, \\ $D_3(x_1,x_2,\vec 0_\perp)$, for the models I (lower panel) and II (upper panel). \label{Fig:dpdf}}
\end{figure}

 Our results for the DPDF calculated for $\vec \eta_\perp=\vec 0_\perp$ are shown for the two considered models in Fig.~\ref{Fig:dpdf}. For this particular value of transverse momentum $D_3(x_1,x_2,\vec 0_\perp)$ the double distribution is the probability density for finding quarks with momentum fraction $x_1$ and $x_2$.
 In the upper panel it is seen that for model II, a strong suppression of the DPDF is seen for $x_1>0.6$ as for the PDF. The model with $a<0$ gives a slightly more narrow DPDF. Observe the different shapes of the boundaries of the double quark distribution, giving complementary information with respect to the  two-quark  transverse momentum distribution, which is sensitive to the size of the  proton, as we are going to discuss. The boundary  for the higher probability density region for model I has an isosceles triangle shape, while for model II it has an isosceles trapezoid shape. The totally symmetric character of the wave function leads to the symmetry properties of the boundaries. 
 
 The triangular shaped boundary for model I, could be anticipated
 from the peak of largest probability for $x_i\sim 0.3$, being visible for 0.2 and above in Fig.~\ref{Fig:dpdf}, up to the boundary $x_1+x_2=1-x_3$.
 The trapezoid shaped boundary observed for model II can be associated with the strong damping of the PDF above $x\sim 0.6$ seen in Fig.~\ref{Fig:PDFs} and to its peak around $x\sim 0.35$, these two properties compete to provide the form seen in 
the upper panel of Fig.~\ref{Fig:dpdf}. Model II corresponds to an excited state and the nodes appearing in the vertex function for $x$ around 0.2 and 0.6 provides such peculiar boundary form. What is noticeable is the sensitivity of the double PDF to the detail of the vertex function, while the valence transverse distributions are essentially sensitive to the size of the three quark configuration. Therefore, it is quite interesting to see that radially excited states have its particular imprints on the double quark distribution, as well as on the PDF.

\subsection{ Transverse momentum densities} 
\label{sec5c}

The single quark transverse  momentum distribution 
 in the forward limit~\cite{Lorce_2011} and integrated in the longitudinal momentum 
is
associated with the probability density to find a quark with momentum $k_{\perp}$:
 \begin{equation}
 \label{TMDq}
 \begin{aligned}
 L&_1(k_{1\perp}) = \frac{k_{1\perp}}{(2\pi)^6}\int_{0}^1 dx_1 \int_0^{1-x_1}dx_2 \int_0^1 dx_3 \\
 & \times  \delta(1 - x_1 - x_2 - x_3) \int_{0}^{2\pi}d\theta_1 \int d^2 k_{2\perp}d^2 k_{3\perp}\\
 &\times\delta(\vec{k}_{1\perp} + \vec{k}_{2\perp} + \vec{k}_{3\perp})|\Psi_3(x_1,\vec{k}_{1\perp}, x_2, \vec{k}_{2\perp}, x_3, \vec{k}_{3\perp})|^2\\
 & \times \,|\Psi_3(x_1,\vec{k}_{1\perp}, x_2, \vec{k}_{2\perp}, x_3, \vec{k}_{3\perp})|^2\, ,
 \end{aligned}
 \end{equation} 
and the two-quark one reads
\begin{small}
\begin{equation}
\label{TMD2q}
\begin{aligned}
 &L_2(k_{1\perp}, k_{2\perp}) = \frac{k_{1\perp}k_{2\perp}}{(2\pi)^6}\int_{0}^1 dx_1 \int_0^{1-x_1}dx_2\int_0^1 dx_3 \\
 &\times  \delta(1 - x_1 - x_2 - x_3)  \int d^2 k_{3\perp}\delta(\vec{k}_{1\perp} + \vec{k}_{2\perp} + \vec{k}_{3\perp})\\
 & \times \int_{0}^{2\pi}d\theta_1
 \int_{0}^{2\pi}d\theta_2|\Psi_3(x_1,\vec{k}_{1\perp}, x_2, \vec{k}_{2\perp}, x_3, \vec{k}_{3\perp})|^2.
 \end{aligned}
 \end{equation} 
\end{small}

\begin{figure}[!htbp]
  \centering
  \includegraphics[scale=0.6]{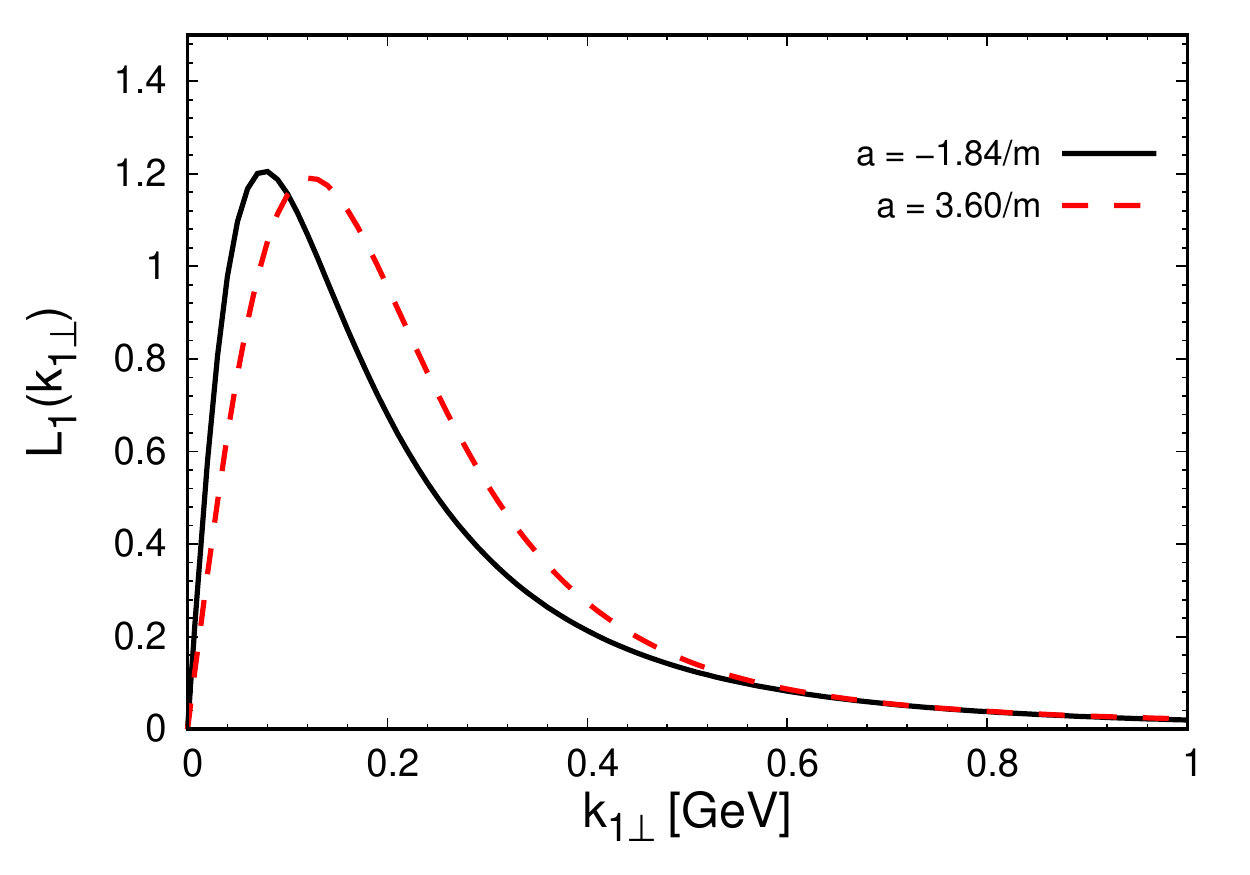}
 \vspace{-0.5cm}
  \caption{One-quark transverse  momentum density vs $k_{1\perp}$ for the models I $(a=-1.84/m)$
  and II $(a=3.60/m)$. \label{Fig:tmd_1q}}
\end{figure}

\begin{figure}[!bhtp]
  \centering
 \includegraphics[scale=0.8]{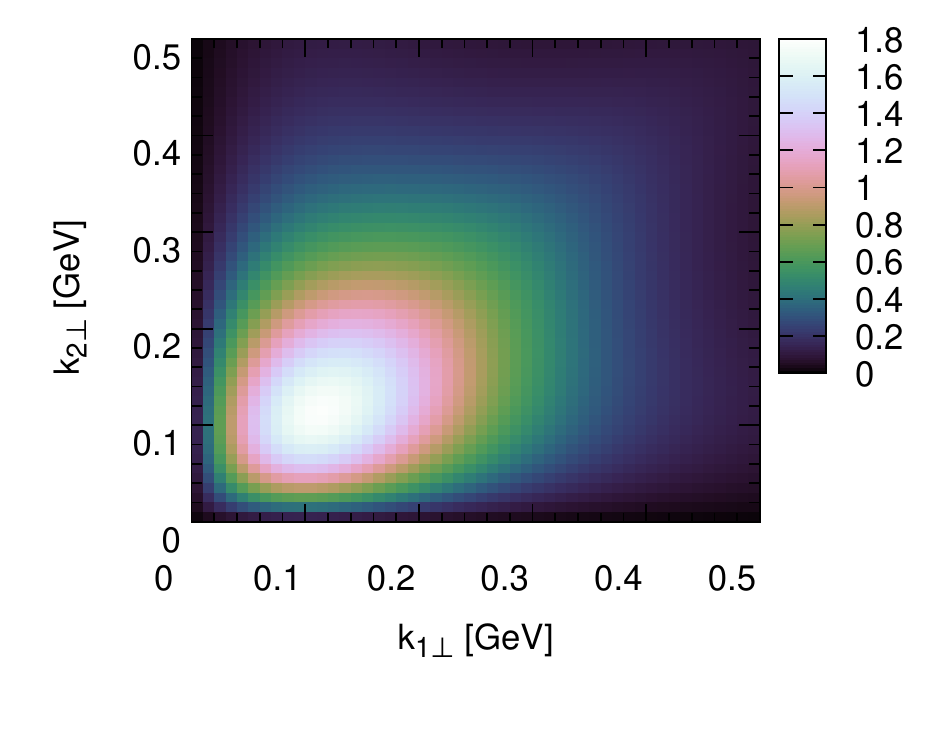}
 \includegraphics[scale=0.8]{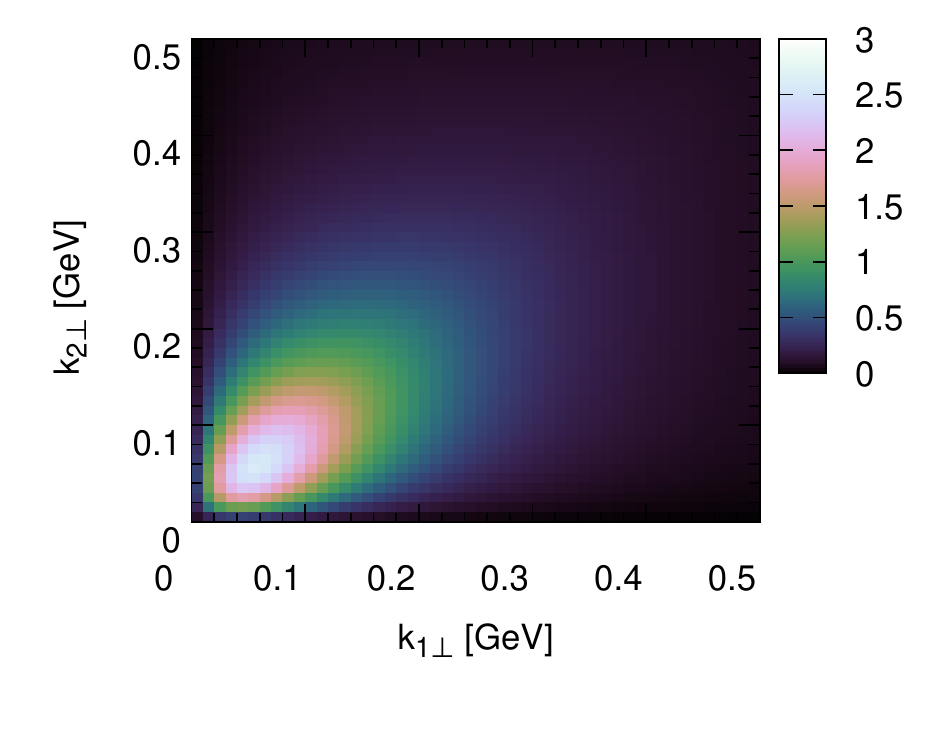}
 \vspace{-1cm}
  \caption{Two-quark transverse  momentum density versus $k_{1\perp}$ and $k_{2\perp}$,  for the model I (lower panel) and II (upper panel). \label{Fig:tmd_2q}}
\end{figure}

In Fig.~\ref{Fig:tmd_1q} the single quark transverse momentum  density from Eq.~\eqref{TMDq} is shown for models I and II. As expected, for model I with $a=-1.84/m$, the momentum distribution is narrower than for $a=3.6/m$, as the radius for the former case is larger. The peak of the momentum distribution is about 0.08~GeV for model I and 0.12~GeV for model II, reflecting the larger size of the proton in model I compared to model II.

In Fig.~\ref{Fig:tmd_2q} the two-quark  transverse momentum density is shown for the model I (lower panel) and model II (upper panel). The more compact configuration of model II is  reflected in the wider distribution, and the probability density peak is consistent with  Fig.~\ref{Fig:tmd_1q} to what was observed for the quark distribution.

\section{ Summary} 
\label{secsummary}

We  study the
Ioffe-time image, non-polarized longitudinal and transverse momentum distributions, and the double momentum distribution of the proton relying on a dynamical constituent quark light-front three-body model.  The dynamics is based on the prevalence of the scalar diquark channel in the quark-quark interaction. We assume a minimal structure of quark-quark contact interaction, resembling the Nambu-Jona-Lasinio model, with the simplified assumption of factorization of the spin degree of freedom, and focusing on the totally symmetric momentum component of the light-front wave function. We study two possibilities, namely a diquark in a bound  or a virtual state.

The three-body light-front Faddeev Bethe-Salpeter equations for the valence state was solved in the presence of  virtual (model I) or bound (model II) diquark states,  for positive and negative scattering lengths, respectively. The contact interaction allows to simplify the integral equations for the Faddeev components of the vertex function, which depend only on  the spectator quark longitudinal momentum fraction and transverse momentum. We have not used a momentum cut-off in the model as originally introduced in~\cite{Frederico92} and adopted the no-cutoff version~\cite{Carbonell03}. This simplified dynamical model allowed to investigate only non-polarized quantities.

The adopted dynamical light-front model has two parameters: the constituent quark mass and scattering length. To determine these parameters the proton mass was fixed to its experimental value, and the binding energy as well as the scattering length were varied to have a qualitative fit of the Dirac form up to about 1~GeV$^2$. It reproduces the Dirac form factor radius  somewhat close to 1~fm, having the case of bound diquark a more compact configuration than the case with the virtual diquark state. These two possibilities produces quite different proton properties for the model with no cut-off. The bound diquark produces a deep three-quark non-physical state $(M^2<0)$, and the one associated with the proton is in this case an excited state, where the vertex function has nodes in the $x$  dependence, while for the virtual diquark the nucleon is the ground state of the model. These two distinct natures of the valence state for model I and II allowed to study their observable consequences in the different momentum distributions and image. 

Specifically, we computed several proton non-\\polarized quantities for the models (I) and (II): 
\begin{enumerate}[label=(\roman*)]
\item  the distribution amplitude (DA); it corresponds to the probability amplitude to find the quarks with given momentum fraction at the same transverse position. The model I has a narrow distribution on the $(x_1,x_2)$ plane when compared to model II. 
\item  the Ioffe-time image; it corresponds to the \\ Fourier-transform of the DA in the Ioffe time, and gives the probability density of finding quarks along the light-like direction for quarks at the same transverse position. The quarks tend to have close  Ioffe-time positions, and exhibit characteristic  oscillations reflecting the size and mass  of the system, besides  the symmetry of the configuration space wave function.
\item  the quark distribution function; it corresponds to the probability density to find a quark with a given momentum fraction at the nucleon scale, peaked around $x\sim 0.3$, but distinguishing model I and II, the last one having nodes in the spectator function, and presenting a more localized distribution.
\item the  double quark distribution function; for $\vec \eta_\perp=0$;  it corresponds to the probability density to find the quarks with  given momentum fractions at the nucleon scale. The shape of the boundary with large probability to observe the momentum fractions distinguish model I  and II, with an isosceles triangular and trapezoid shapes, for respectively, the ground-state and excited-state configurations.
\item  the 
  single quark transverse momentum  density; is associated with the probability density to find a quark with a given transverse momentum and mainly sensitive to the size of the three-quark configuration and found peaked around 0.1~GeV.
\item   the double quark transverse momentum  density;  is associated with the probability density to find the quarks with a given transverse momentum and, again, mainly sensitive to the size of the three-quark configuration, a smaller region in momentum is found for the  larger size of the proton for model I compared to the more compact configuration of the quarks in model II. 
 \end{enumerate}

Future challenges for improving the nucleon effective model to be taken: the  computation of the Bethe-Salpeter amplitude in the four-dimensional Minkowski space, which includes an infinite number of Fock-components, the introduction of a cut-off and the spin degree of freedom, which we expect will provide more insights into the nucleon structure.
 
\begin{acknowledgments}
This work is a part of the project INCT-FNA proc. No. 464898/2014-5.
This study was financed in part by Conselho  Nacional de Desenvolvimento Cient\'{i}fico e  Tecnol\'{o}gico (CNPq) under the grant 308486/2015-3 (TF). E.Y.~thanks for the financial
support of the grants \#2016/25143-7 and \#2018/21758-2 from FAPESP.  We  thank the FAPESP Thematic Projects grants   \#13/26258-4 and \#17/05660-0.  
\end{acknowledgments}

\newpage

\appendix
\section{Derivation of the LF wave function and vertex equation}
\label{App:Derivation}

The derivations made in this Appendix are based on the LF projection technique of the BS equation and corresponding amplitude  based on the Quasi-Potential expansion developed in~\cite{Sales:1999ec} for the two-boson problem and in Refs.~\cite{Marinho:2007zz,Frederico:2010zh,Guimaraes:2014kor} for the three-boson case. For the sake of completeness, we sketch here the main steps in deriving the valence wave function~\eqref{Eq:BS_wf} and the associated Faddeev equation for the  vertex function~\eqref{Eq:3b_LF}.

The starting point is the three-boson  BS amplitude,  which is defined as: 
\begin{equation} \label{Phi0} \Psi_M(y_1,y_2,y_3;p)=\langle 0 \left| T(\varphi (y_1)\varphi (y_2)\varphi (y_3))\right| p\rangle \, ,\end{equation}
where $y_i$ is the space-time position of  particle $i$, $\varphi(y)$ is the bosonic field operator and $p$ the total momentum.

The valence LF wave function comes from the
projection of the BS amplitude onto the null-plane: 
\begin{multline} \label{bs7b}
\psi_3 (\vec{k}_{1\perp},x_1;\vec{k}_{2\perp},x_2;\vec{k}_{3\perp},x_3)=(p^+)^2 \, (x_1 x_2 x_3)^\frac12
\\ \times\chi_3 (\vec{k}_{1\perp},x_1;\vec{k}_{2\perp},x_2;\vec{k}_{3\perp},x_3)\, ,
\end{multline}
with
\begin{multline} \label{bs7c}
\chi_3 (\vec{k}_{1\perp},x_1;...)=
\int dk^{-}_{1}\, dk^{-}_{2}\, 
\Phi_M(k_1,k_2,k_3;p)\, ,
\end{multline}
where $\Phi_M$ is the momentum representation of the Minkowski space BS amplitude $\Psi_M$. We have introduced the auxiliary LF amplitude $\chi_3$ for the  convenience of the derivations done in what follows.

The elimination of the relative
LF time for the three-body BS equation and associated amplitude requires an integration  over two independent momenta $k^-$, due to four-momentum conservation,
and we introduce the following operation for a quantity defined in Minkowski space:
\begin{eqnarray}
|A&:=&\int dk_1^-dk_2^-\langle k_1^-k_2^-|A,\nonumber \\
A|&:=&\int dk_1^-dk_2^-A|k_1^-k_2^-\rangle,
\label{bardef3}
\end{eqnarray}
with $A$ being a matrix element of an operator that has matrix elements
which are functions of two independent momenta after the center of mass motion
is factorized.
With the above notation the LF wave function is:
\begin{equation}
|\chi_3\rangle=\big||\Phi_M\rangle=\big|G_0|\Gamma_M\rangle
\end{equation}
where the BS equation for the vertex function is
\begin{eqnarray}
  |\Gamma_M\rangle&=& V  G_0 |\Gamma_M\rangle.
\end{eqnarray}
Explicitly the  three-particle free Green's function is given by
\begin{small}
\begin{multline}
\langle k_1^-,k_2^- |G_0|k_1^{\prime -},k_2^{\prime -}
\rangle=\frac{-i}{(2\pi)^2}\frac{\delta(k_1^--k_1^{\prime -})}
{\hat{k}_1^+\hat{k}_2^+(p^+-{\hat{k}_1}^+-\hat{k}_2^+)}\\ \times \frac{\delta(k_2^--k_2^{\prime -})}
{
(k_1^--\hat{k}_{1 on}^-)(k_2^--\hat{k}_{2
on}^-)(p^--k_1^--k_2^--\hat{k}_{3on}^-)}~,
\end{multline}
\end{small}
where the hat means operator character and the on-minus-shell momentum $ \hat{k}^-_{ion}=(\hat{\vec k}^2_{i\perp}+m^2_i)/\hat{k}_i^+$.
The momentum conservation applies to the kinematical components of the momentum of particle 3, such that $\hat k^+_3=p^+-\hat{k}^+_1-\hat{k}^+_2$  and the analogous expression for the transverse components. 
By performing the LF projection using Eq.~(\ref{bardef3}), the free LF
Green's function becomes
    $g_0 = |G_0|,$
being the free light-front resolvent, explicitly written as:
\begin{multline}
g_0(\underline{k}_1,\underline{k}_2)
=\frac{i\theta(p^+-k_1^+-k_2^+)\theta(k_1^+)\theta(k_2^+)}{k_1^+k_2^+(K^+-{k_1}^+-k_2^+)} \\
\times\frac{1}{p^--k_{1 on}^--k_{2 on}^--(p-k_1-k_2)_{on}^-}
~,\label{propfl3}
\end{multline}
where $\underline{k}_i\equiv\{ 
k^+_i,\vec k_{i\perp}\}$.

For the auxiliary Green's function $\tilde{G}_0$ one makes the choice
\begin{equation}
\label{Eq:G_tilde_0}
    \widetilde{G}_0 = G_0|g^{-1}_0|G_0,
\end{equation}
and with that the four-dimensional BS equation for the vertex function is also a solution of:
\begin{equation}
  |\Gamma_M\rangle= W  \widetilde G_0 |\Gamma_M\rangle=W G_0|g^{-1}_0|G_0|\Gamma_M\rangle\, ,
\end{equation}
where the quasi-potential $W$ is given by
\begin{equation}
\label{Eq:W}
W = V + V \Delta_0 W
\end{equation}
with
$ \Delta_0 = G_0 - \widetilde{G}_0\, .$

The LF auxiliary amplitude turns out to be:
\begin{equation}
  |\chi_3\rangle=\big|G_0   |\Gamma_M\rangle=\big |G_0 W G_0\big|g^{-1}_0|G_0\big|\Gamma_M\rangle\, ,
\end{equation}
and introducing the LF interaction
\begin{equation}
\label{Eq:w_L0}
    w = g_0^{-1}|G_0 W G_0| g_0^{-1}\, ,
\end{equation}
and substituting in the expressions for $|\chi_3\rangle$:
\begin{equation}
  |\chi_3\rangle=g_0\,w|G_0\big|\Gamma_M\rangle =
  g_0\,w|\chi_3\rangle \, ,
\end{equation}
where the LF vertex function is a solution of
\begin{equation}
    |\Gamma_{LF}\rangle =w\, g_0|\Gamma_{LF}\rangle\, .
\end{equation}

We introduce in  what follows the Faddeev decomposition to solve the above equation for the vertex function.
The potential is built from the two-body ones as
\begin{equation}
\label{{Eq:V}}
    V = \sum_{i=1}^3 V_i,\quad\text{and}\quad V_i = V_{(2)i}S^{-1}_i
\end{equation}
where $S_i$ is the propagator of the particle $i$ and $ V_{(2)i}$ is the interaction between the particles $j$ and $k$.

The Faddeev decomposition of $W$ reads
\begin{equation}
\label{Eq:Faddeev_W}
    W = \sum_{i} W_i \quad\text{and}\quad w=\sum_{i}w_i,
\end{equation}
where
$
    w_i = g_0^{-1}|G_0 W_i G_0| g_0^{-1}\, ,$
and formally the LF three-body wave function is:
\begin{equation}
  |\chi_3\rangle =
  g_0\,\sum_i|v_i\rangle
  \quad\text{with}\quad |v_i\rangle=w_i|\chi_3\rangle\, ,
\end{equation}
where $|v_i\rangle$ are the Faddeev components of the LF vertex function, namely $|\Gamma_{LF}\rangle=\sum_i|v_i\rangle$.

 In terms of the pairwise interaction, $V_i$, one has
\begin{equation}
\label{Eq:W_exp}
\begin{aligned}
    W_i &= V_i +V_i \Delta_0(V_i + V_j + V_k) \\
    & + V_i\Delta_0(V_i + V_j + V_k)\Delta_0(V_i + V_j + V_k) + \cdots 
    \end{aligned}
\end{equation}

At the lowest (LO) order the effective potential  is
\begin{equation}
\label{Eq:w_L1}
    w^{\text{LO}}_i = g_0^{-1}|G_0 V_i G_0| g_0^{-1} \, ,
\end{equation}
and the  Faddeev equations for  the components of the vertex
read 
\begin{eqnarray}
    \label{Eq:v_eq_2}
  |v^{\text{LO}}_i \rangle&=&  t^{\text{LO}}_i g_0 (|v^{\text{LO}}_j\rangle + |v^{\text{LO}}_k\rangle)\, ,  
\end{eqnarray}
with the LF T-matrix being a solution of
\begin{equation}\label{Eq:tLO}
    t^{\text{LO}}_i = w^{\text{LO}}_i+w^{\text{LO}}_i g_0\,t^{\text{LO}}_i\,.
\end{equation}
Furthermore, it  should be noted that the LF resolvent and  potential are immersed in the three-body system.

For the contact interaction the
matrix element of the potential $V_i$ is
\begin{equation}
\label{Eq:V_me}
    \langle k_j, k_k |V_i|k'_j, k'_k \rangle = \lambda (2\pi)^2 \delta(k_i - k'_i)(k_i^2 - m^2)\, .  
\end{equation}
By introducing it in Eq.~\eqref{Eq:w_L1}, solving the LF T-matrix equation~\eqref{Eq:tLO}, and also taken into that the two-body system is immersed in the three-body one, it is found that: 
\begin{equation}
\label{Eq:me}
    \Bigl\langle  \underline k_j, \underline k_k \Bigl|t^{\text{LO}}_i \Bigr|  \underline k'_i, \underline k'_j\Bigr\rangle =
    -i\mathcal{F}(M^2_{jk}) k^+_i \delta(\underline k_i - \underline k'_i),
\end{equation}
where $M^2_{jk}=(p - k_{i,on})^2$ and momentum conservation implies that $\underline p=\underline k_i+\underline k_j+\underline k_k=\underline k'_i+\underline k'_j+\underline k'_k\, .$

Furthermore, for the zero-range interaction we have that $V_i\Delta_0 V_i = 0$ and it then follows that $W_i = V_i$ at the valence order.
We thus obtain
\begin{small}
\begin{equation}
\label{Eq:v_eq_3}
    v^{\text{LO}}_i(\underline k_i) = -2i \mathcal{F}(M^2_{jk}) \int d\underline k'_j k^+_i g_0(\underline k'_i, \underline k'_j)v^{\text{LO}}_j(\underline k'_j)\, ,
\end{equation}
\end{small}
where the measure is
$d\underline k\equiv\frac{dk^+d^2k_\perp}{2(2\pi)^3}$ and 
the factor of two comes from the symmetrization of the total vertex function with respect to the exchange of the bosons, which also
implies that $$v^{\text{LO}}_i(\underline k_i)=v^{\text{LO}}_j(\underline k_j)=v^{\text{LO}}_k(\underline k_k)\,. $$ Finally, identifying $\Gamma(x_i,k_{i\perp})\equiv v^{\text{LO}}_i(\underline k_i)$, with $x_i=k^+_i/K^+$ one finds Eq.~\eqref{Eq:3b_LF} from Eq.~\eqref{Eq:v_eq_3}.

\section{Numerical methods}
\label{App:NM}

 In the present work the homogeneous integral equation \eqref{Eq:3b_LF} was solved for given values of $a$ and $M_3 = M_N/m$ by expanding the vertex function $\Gamma(x,k_{\perp})$ in a bicubic basis, on the domain $\Omega=I_{k_\perp}\times I_{x}=[0,\infty]\times[0,1]$, of the form
\begin{equation}
\Gamma(x,k_\perp) = \sum_{i=0}^{2N_{k_\perp}}\sum_{j=0}^{2N_x - 1}A_{ij}S_i(k_\perp)S_j(x).
\end{equation}.
In the calculations the intervals for $k_\perp$ and $x$ were partitioned into $N_{k_\perp}$ and $N_x$ subintervals, respectively.

The equation \eqref{Eq:3b_LF} can then be turned into a generalized eigenvalue problem of the form
\begin{equation}
\label{Eq:ev_eq}
    \sum_{i'j'}P_{ij i'j'}A_{i'j'} = \lambda(M_3)\sum_{i'j'}V_{iji'j'}A_{i'j'},  
\end{equation}
where $P_{iji'j'}=S_{i'}(k^{(i)}_\perp)S_{j'}(x^{(j)})$ and $V_{iji'j'}$ is  the \\ right-hand side of \eqref{Eq:3b_LF} with $\Gamma$ replaced by \\ $S_{i'}(k^{(i)}_\perp)S_{j'}(x^{(j)})$. The value of the three-body mass $M_3$ were  found by  iteratively solving the non-linear equation $\lambda(M_3) = 1$ and the corresponding coefficients from the solution of \eqref{Eq:ev_eq}. The obtained solution for the vertex function was subsequently normalized by imposing the condition $F(0)=1$, where $F(Q^2)$ is the valence LF Dirac form factor which is discussed in Sec.~\ref{sec4}.

\newpage 



%

\end{document}